\documentclass[]{JFM-FLM_Au}
\usepackage{amsmath,mathtools}
\usepackage{physics}
\usepackage{microtype}
\usepackage{subcaption}      % for (a), (b), (c) subfigures
\setcitestyle{round}         % natbib is already loaded by the class
\usepackage{bm}               % bold math symbols
\hypersetup{hidelinks}
\newcommand{\ii}{\mathrm{i}}

\newcommand{\vect}[1]{\bm{#1}}
\graphicspath{{figuresdh/}}
\numberwithin{equation}{section}

\lefttitle{Traore and Halpern}
\righttitle{Journal of Fluid Mechanics}

\title{Stability and nonlinear dynamics of three-layer viscous films inside a vertical cylindrical tube}

\author{Awa Traore\aff{1} \and David Halpern\aff{1}}

\affiliation{\aff{1}Department of Mathematics, University of Alabama,
  Tuscaloosa, AL 35487, USA}

\corresau{David Halpern, \email{dhalpern@ua.edu}}

\begin{document}
\maketitle

\begin{abstract}
  We investigate the dynamics and stability of three immiscible viscous liquid  
  layers coating the interior of a vertical cylindrical tube, a configuration
  relevant to stratified core--annular transport processes. A long-wave asymptotic analysis yields a coupled system of nonlinear
  evolution equations governing the motion of the three interfaces. Linear stability analysis predicts a persistent long-wave instability, the capillary (Rayleigh--Plateau) instability of the air--core interface, together
  with secondary finite-wavenumber instability bands that emerge from interfacial   coupling in certain parameter regimes. These  stability 
  characteristics depend sensitively on the layer   thicknesses, viscosity ratios, and surface tension parameters, and include mode-switching associated with competing maxima in the dispersion   relation.
  
  Nonlinear simulations reveal three distinct dynamical outcomes: saturation to
  finite-amplitude travelling waves, air-core closure through plug formation,
  and rupture of the intermediate liquid layer while the air core remains open.
  The intermediate-layer rupture mechanism is unique to the three-layer
  configuration which has no analogue in one- or two-interface cylindrical film flows. Numerical continuation is used to compute branches of
  travelling-wave solutions and their associated limit points. Comparison with time-dependent simulations shows that travelling-wave branches successfully predict the transition from saturated waves to plug formation, but do not capture the distinct rupture mechanism associated with collapse of the intermediate layer.
  \end{abstract}

\setcounter{section}{0}   % This forces the next \section to be numbered 2
\setcounter{equation}{0}  % Reset equation counter to start at (2.1)
\setcounter{figure}{0}    % Adjust if your first figure in this section is Figure 3

\section{Introduction}
Multi-layer viscous films are a defining feature of systems ranging from the micro-scale mechanics 
of human lungs to the macro-scale transport of multiphase fluids in industrial pipelines. Recent 
reviews have highlighted the central role of interfacial dynamics in both pulmonary airway function 
and stratified industrial flows \citep{romano2026human,Peng2025}. In human bronchioles, the 
airway surface liquid is described by a low-viscosity periciliary layer beneath an overlying mucus 
layer, together with a surfactant film at the air–liquid interface \citep{Heil2008}. Analogous 
multilayer configurations arise in water-lubricated core–annular transport of heavy crude oils, in 
which a viscous oil core is conveyed within a lubricating annulus, and in multilayer co-extrusion and 
coating processes used to manufacture fibres, films, and encapsulated microstructures 
\citep{Joseph1997,Lamnawar2013,Peng2025}. In these systems, the dynamics are governed by the 
interaction of multiple deformable interfaces, whose evolution controls flow stability and transport. 
In both settings, sufficiently large interfacial deformations may obstruct the flow by producing 
airway closure or pipeline blockage, motivating the study of coupled interfacial instabilities.

Beyond any single application, the three-layer configuration is the minimal
setting in which two deformable internal interfaces interact within a confining
geometry, making it a natural framework for asking whether the
multiplicity of interfaces produces qualitatively new instability and rupture
mechanisms rather than merely quantitative modifications of the two-layer
problem. In planar and inclined geometries this question has received some attention. Early studies 
established that viscosity stratification can destabilise interfaces even in the absence of inertia 
\citep{Kao1965a,Kao1965b,Loewenherz1989,Chen1993}. Building on this foundation, \citet{Jiang2005} showed that viscosity-stratified three-layer flow down an inclined wall can support inertialess instabilities with growth rates orders of magnitude larger than those of comparable two-layer flows, while the nonlinear evolution of the interfaces remains well described by linear theory even at large amplitude, a picture supported by their own experimental observations of three-layer gelatine flows. \citet{Henry2014} complement this with a systematic experimental validation of the single-layer thickness and velocity approximation across two- and three-layer falling films, confirming that the base-flow description underlying such stability analyses remains robust even when the layers differ substantially in viscosity. Together, such studies confirm that additional interfaces enrich the instability structure while the underlying base flow remains well characterised experimentally, but the planar setting lacks the azimuthal curvature responsible for the capillary instability and plug formation that are central to cylindrical confinement. Despite this, the
stability of \emph{three}-layer immiscible films inside a vertical cylindrical
tube has received little attention to date. The study of capillary instability in cylindrical geometries 
dates back to
\citet{Rayleigh1878}, who analysed the instability of liquid jets, and
\citet{Tomotika1935}, who considered a viscous liquid thread surrounded by
another viscous fluid. \citet{Goren1962} subsequently investigated the instability of annular liquid coatings, including films coating the inside of a small tube.
Building on these classical studies, \citet{Hammond1983} analysed the nonlinear evolution of a thin 
liquid film coating the inner surface of a cylindrical tube using a long-wave approximation. Later, \citet{Gauglitz1988,Gauglitz1990} retained higher-order curvature contributions in the evolution equation and demonstrated conditions under which finite-time plug formation can occur. 
More recently, two-layer systems have received increasing attention 
\citep{Ogrosky2021,Erken_Romanò_Grotberg_Muradoglu_2022}. In particular, 
\citet{Erken_Romanò_Grotberg_Muradoglu_2022} investigated capillary instability and airway 
closure in a two-layer annular film model motivated by pulmonary airway occlusion.The additional 
viscosity ratios and interfaces introduced by a third
layer, however, can give rise to coupled interfacial modes and nonlinear rupture
mechanisms that are absent in both single- and two-layer systems. In this work
we address this gap by deriving a long-wave evolution model for a three-layer
concentric configuration and conducting systematic linear and nonlinear
stability analyses across the parameter space of viscosity ratios, surface
tension ratios, and layer thicknesses. Preliminary long-wave modelling and nonlinear simulations for 
related three-layer cylindrical configurations were reported in the doctoral thesis of 
\citet{Traore2024}. The present paper extends that work through the inclusion of 
arbitrary-wavenumber linear stability analysis, nonlinear travelling-wave computations, numerical 
continuation, and a systematic comparison between travelling-wave structure and time-dependent 
rupture dynamics.

The governing evolution equations are derived in
Section~\ref{sec:model_eqns} from the axisymmetric Navier--Stokes
equations using the outer-layer viscosity, thickness, and surface tension
as reference scales. In Sections~\ref{sec:aw} and~\ref{sec:lw_linear}
we investigate the linear stability problem and determine how the
instability characteristics depend on the layer thicknesses, viscosity
ratios, and surface tension ratios. The analysis characterises the instability bands and determines how they depend on the layer thicknesses, viscosity ratios, and surface tension ratios. In Section~\ref{sec:nonlinear_evolution} we solve the nonlinear evolution equations numerically and identify three qualitatively distinct outcomes: saturation to finite-amplitude travelling waves, plug formation through closure of the air core, and rupture of the intermediate liquid layer. 
We then compute nonlinear travelling-wave solutions and employ numerical continuation to determine how the solution branches and their limit points vary with viscosity ratio and mean layer thickness.
Comparison with time-dependent simulations demonstrates that the travelling-wave solutions provide useful insight into the onset and character of strongly nonlinear deformation, while revealing how various rupture mechanisms dominate in different parameter regimes. Conclusions and directions for future work are presented in Section~\ref{sec:conclusion}.

\section{Model Equations}\label{sec:model_eqns}

\subsection{Problem setup and geometry}

\begin{figure}
  \centering
  \includegraphics[width=0.95\textwidth]{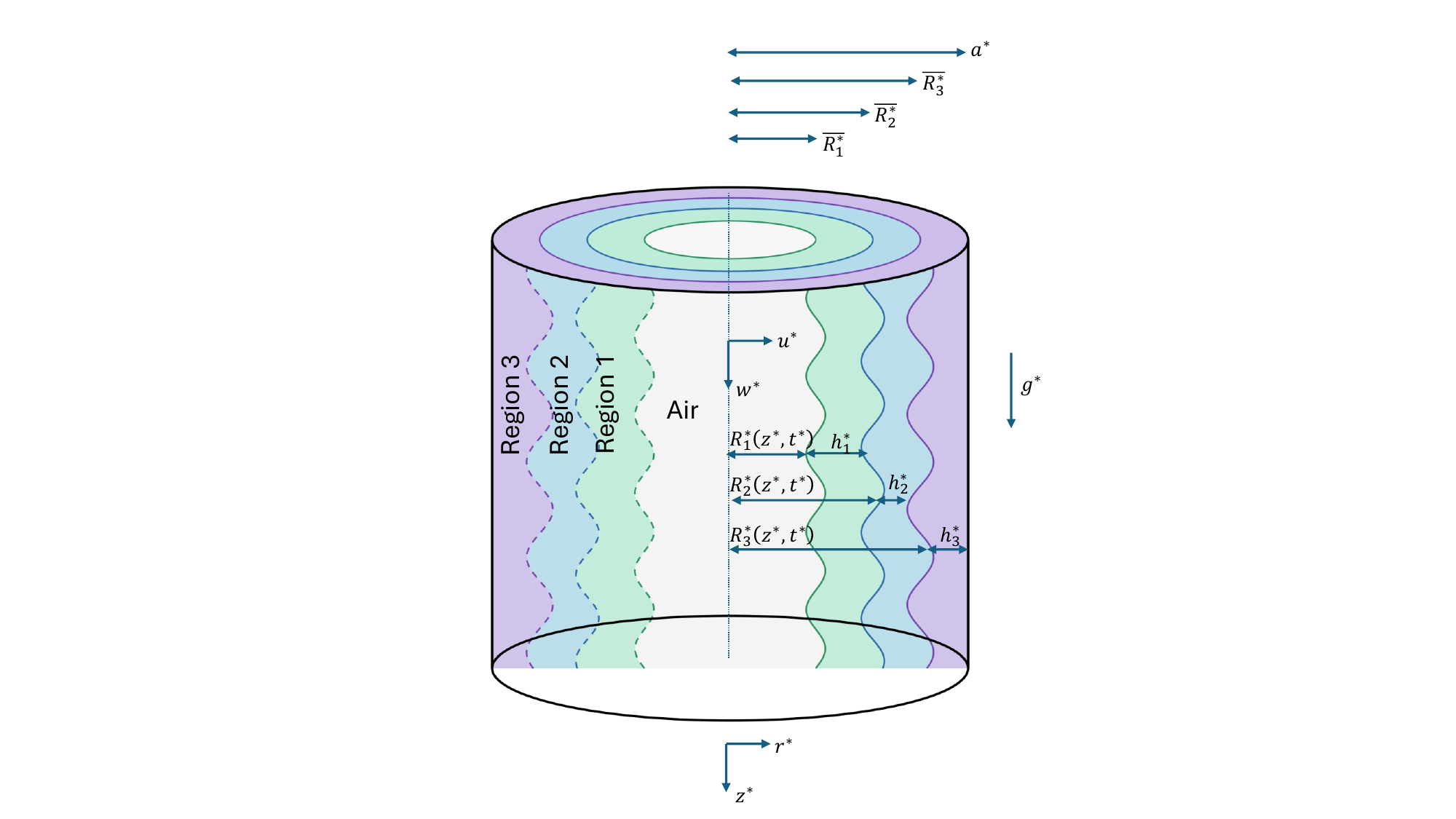}
  \caption{Falling three-layer fluid system inside a cylinder with concentric annular fluid layers and a passive air core.}
  \label{fig:geometryfig}
\end{figure}

We consider the axisymmetric flow of three immiscible fluids down the interior of a rigid vertical tube of radius $a^*$. The fluids form concentric annular layers surrounding a passive air core, as illustrated in  figure~\ref{fig:geometryfig}. Cylindrical coordinates $(r^*,z^*,\theta^*)$ are used, with $z^*$ directed downward (the direction of gravity), $r^*$ the radial coordinate, and no azimuthal dependence. Starred quantities denote dimensional variables, unstarred quantities dimensionless variables, and overbars unperturbed base-state values. 
The domain is divided by three interfaces at $r^*=R_1^*, R_2^*, R_3^*$ (all functions of $z^*,t^*$): the air core lies in $0<r^*<R_1^*$, and fluid layers 1, 2, and 3 fill the successive annular regions $R_1^*<r^*<R_2^*$, $R_2^*<r^*<R_3^*$, and $R_3^*<r^*<a^*$, respectively. Each liquid layer $j$ ($j=1,2,3$) has constant viscosity $\mu_j^*$ and surface tension $\sigma_j^*$ at its outer boundary. The layer thickness is denoted by $h_j^*(z^*,t^*)$, with unperturbed value $\bar h_j^*$. The three fluids are taken to have the same density $\rho^*$, 
and the air core is dynamically passive. In the unperturbed state,
\[
  \bar h_1^*+\bar h_2^*+\bar h_3^*=a^*-\bar R_1^*.
\]

\subsection{Governing equations}
The motion of the viscous fluids in each layer is governed by the axisymmetric Navier--Stokes 
equations together with the continuity equation. In dimensional form, these are
\begin{align}
  \rho_i^*\left(
  u_{i,t^*}^* + u_i^* u_{i,r^*}^* + w_i^* u_{i,z^*}^*
  \right)
  &= -p_{i,r^*}^*
    + \mu_i^*\left(
    \frac{1}{r^*}(r^* u_{i,r^*}^*)_{r^*}
    + u_{i,z^*z^*}^*
    - \frac{u_i^*}{r^{*2}}
    \right),
    \label{eq:NS_r}
  \\
  \rho_i^*\left(
  w_{i,t^*}^* + u_i^* w_{i,r^*}^* + w_i^* w_{i,z^*}^*
  \right)
  &= -p_{i,z^*}^*
    + \mu_i^*\left(
    \frac{1}{r^*}(r^* w_{i,r^*}^*)_{r^*}
    + w_{i,z^*z^*}^*
    \right)
    + \rho_i^* g^*,
    \label{eq:NS_z}
  \\
  \frac{1}{r^*}(r^* u_i^*)_{r^*} + w_{i,z^*}^* &= 0,
                                                 \label{eq:continuity}
\end{align}
for $i=1,2,3$, where $(u_i^*,w_i^*)$ are the radial and axial velocity components and $p_i^*$ is 
the pressure in layer $i$. Subscripts after the comma denote partial differentiation.

\subsection{Boundary conditions}

At the tube wall $r^*=a^*$, no-slip and no-penetration conditions are imposed:
\begin{equation}
  u_3^* = 0, \qquad w_3^* = 0. \label{eq:wall_bc}
\end{equation}

At the internal interfaces $r^* = R_j^*(z^*,t^*)$, $j=2,3$, we apply continuity of velocity, tangential stress balance, normal stress balance, and the kinematic condition. Below, we use the notation $[\cdot]_{i=j-1}^{i=j}$ to denote the jump in a quantity across the interface, i.e.\ the value in layer $j$ minus that in layer $j-1$.
\begin{equation}
  u_{j-1}^* = u_j^*, \qquad w_{j-1}^* = w_j^*, \label{eq:vel_cont}
\end{equation}
\begin{equation}
  \begin{aligned}
    \bigl[\, &(1-(R_{j,z^*}^*)^2)
               \bigl(w_{i,r^*}^* + u_{i,z^*}^*\bigr)
    \\
             &\quad + 2R_{j,z^*}^*
               \bigl(u_{i,r^*}^* - w_{i,z^*}^*\bigr)
               \,\bigr]_{i=j-1}^{i=j}
               = 0,
  \end{aligned}
  \label{eq:tangl_stress}
\end{equation}
\begin{equation}
  \begin{aligned}
    \bigl[\, &2\mu_i^*\bigl(u_{i,r^*}^*
               + w_{i,z^*}^*(R_{j,z^*}^*)^2
               - R_{j,z^*}^*(w_{i,r^*}^* + u_{i,z^*}^*)\bigr)
    \\
             &\quad - \bigl(1+(R_{j,z^*}^*)^2\bigr)p_i^*
               \,\bigr]_{i=j-1}^{i=j}
    \\
             &= \sigma_j^*\left(
               \frac{\bigl[1+(R_{j,z^*}^*)^2\bigr]^{1/2}}{R_j^*}
               -
               \frac{R_{j,z^*z^*}^*}{\bigl[1+(R_{j,z^*}^*)^2\bigr]^{1/2}}
               \right),
  \end{aligned}
  \label{eq:norm_stress}
\end{equation}
\begin{equation}
  u_j^*(R_j^*,z^*,t^*)
  =
  \frac{\partial R_j^*}{\partial t^*}
  +
  w_j^*(R_j^*,z^*,t^*)\,
  \frac{\partial R_j^*}{\partial z^*}.
  \label{eq:kinematic}
\end{equation}

At the free surface $r^*=R_1^*(z^*,t^*)$, the air viscosity is neglected. The kinematic 
condition, 
vanishing tangential stress, and normal stress balance are
\begin{equation}
  u_1^*(R_1^*,z^*,t^*)
  =
  \frac{\partial R_1^*}{\partial t^*}
  +
  w_1^*(R_1^*,z^*,t^*)\,
  \frac{\partial R_1^*}{\partial z^*},
  \label{eq:kin_free}
\end{equation}
\begin{equation}
  \begin{aligned}
    &(1-(R_{1,z^*}^*)^2)
      \bigl(w_{1,r^*}^* + u_{1,z^*}^*\bigr)
    \\
    &\qquad
      + 2R_{1,z^*}^*
      \bigl(u_{1,r^*}^* - w_{1,z^*}^*\bigr)
      = 0,
  \end{aligned}
  \label{eq:tan_free}
\end{equation}
\begin{equation}
  \begin{aligned}
    &2\mu_1^*
      \Bigl(
      u_{1,r^*}^*
      + w_{1,z^*}^*(R_{1,z^*}^*)^2
      - R_{1,z^*}^*
      \bigl(w_{1,r^*}^* + u_{1,z^*}^*\bigr)
      \Bigr)
    \\
    &\qquad
      - \bigl(1+(R_{1,z^*}^*)^2\bigr)p_1^*
    \\
    &\qquad
      = \sigma_1^*
      \left(
      \frac{\bigl(1+(R_{1,z^*}^*)^2\bigr)^{1/2}}{R_1^*}
      -
      \frac{R_{1,z^*z^*}^*}{\bigl(1+(R_{1,z^*}^*)^2\bigr)^{1/2}}
      \right).
  \end{aligned}
  \label{eq:norm_free}
\end{equation}

\subsection{Scalings }
We introduce, as in \citet{Ogrosky2021}, a characteristic axial length scale $\lambda^*$ associated with the typical wavelength of interfacial disturbances, and scale radial lengths with the unperturbed outer-layer thickness $\bar h_3^*$. The dimensionless parameter
\begin{equation}
  \epsilon=\frac{\bar h_3^*}{\lambda^*}
\end{equation}
represents the long-wave (``thin-film'') aspect ratio. In the
long-wave regime considered later, we assume $\epsilon \ll 1$, so that
axial variations occur over length scales much larger than the film
thickness.

The characteristic axial velocity scale is chosen to be the gravitational
drainage velocity in the outer layer,
\begin{equation}
  W^*=\frac{\rho^* g^*(\bar h_3^*)^2}{\mu_3^*}.
\end{equation}
The radial velocity scale then follows from the continuity equation and is
smaller by a factor of $\epsilon$, giving
\begin{equation}
  U^*=\epsilon W^*.
\end{equation}
The nondimensional variables are defined by
\begin{equation}
  r = \frac{r^*}{\bar{h}_3^*}, \quad
  z = \frac{z^*}{\lambda^*}, \quad
  u_i = \frac{u_i^*}{U^*}, \quad
  w_i = \frac{w_i^*}{W^*}, \quad
  t = \frac{t^* W^*}{\lambda^*}, \quad
  p = \frac{p^* \bar{h}_3^*}{\mu_3^* W^*}. \label{eq:nondim}
\end{equation}
As a result of \eqref{eq:nondim}, we introduce the parameters listed below:
\begin{equation}
  \begin{aligned}
    a &= \frac{a^*}{\bar{h}_3^*}, 
    &\qquad
      \bar{h}_j &= \frac{\bar{h}_j^*}{\bar{h}_3^*}, 
    &\qquad
      \bar{R}_j &= \frac{\bar{R}_j^*}{\bar{h}_3^*}, 
    \\[0.5em]
    m_1 &= \frac{\mu_3^*}{\mu_1^*}, 
    &\qquad
      m_2 &= \frac{\mu_3^*}{\mu_2^*}, 
    \\[0.5em]
    \sigma_1 &= \frac{\sigma_3^*}{\sigma_1^*}, 
    &\qquad
      \sigma_2 &= \frac{\sigma_3^*}{\sigma_2^*}, 
    \\[0.5em]
    C &= \frac{\mu_3^* W^*}{\sigma_3^*}, 
    &\qquad
      Re &= \frac{\rho^* W^* \bar{h}_3^*}{\mu_3^*}.
  \end{aligned}
  \label{eq:params}
\end{equation}
Here $m_1,m_2$ are viscosity ratios, $\sigma_1,\sigma_2$ are surface-tension ratios, $C$ is a
capillary number, and $Re$ is a Reynolds number. Below, we also use the parameter $m_3=\frac{\mu_3^*}{\mu_3^*}=1$ for convenience.

\subsection{Basic-state quantities}\label{sec:aw_base}

For the steady concentric base state with interface radii $\bar R_1,\bar R_2,\bar R_3$, the axial base 
velocities take the form
\begin{equation}\label{eq:aw_base_w}
  \bar w_j(r)=-\frac{m_j}{4}r^2+a_j\ln r+b_j,   \qquad j=1,2,3,
\end{equation}
where the constants $a_j$ and $b_j$ are determined by the
interfacial and boundary conditions listed in Appendix~\ref{app:base_coeffs}.
We denote the base velocities at the three interfaces by $\bar w_1=\bar w_1(\bar R_1)$, $\bar 
w_2=\bar w_1(\bar R_2)=\bar w_2(\bar R_2)$, and $\bar w_3=\bar w_2(\bar R_3)=\bar w_3(\bar 
R_3)$. We also write
\[
  \tau_j(r)=\bar w_{j,r}(r), \qquad \tau_j'(r)=\bar w_{j,rr}(r).
\]
The quantities $\tau_j(\bar R_i)$ and $\tau_j'(\bar R_i)$ enter the linearised interfacial conditions given in the next section which describes the linear stability for arbitrary wavelengths.

\section{Arbitrary-wavenumber linear stability formulation}\label{sec:aw}

We consider the linear stability of the steady concentric base state with
constant interface radii
\[
  \bar R_1<\bar R_2<\bar R_3<a.
\]
We restrict attention to the Stokes (inertialess) limit, $Re\ll1$, which is
appropriate for the thin, highly viscous films considered here and in which the
inertial terms in \eqref{eq:NS_r}--\eqref{eq:NS_z} are negligible. Axisymmetric
disturbances with axial wavenumber $k$ are introduced, and the resulting Stokes
equations are linearised about the base state. Eliminating the pressure and
axial velocity leads to a fourth-order equation for the radial velocity in each
layer, together with interfacial and wall conditions that couple the three
layers.

The resulting system yields a dispersion relation between the complex growth
rate $s$ and the wavenumber $k$. The formulation is presented in terms of the
radial velocity amplitudes, which provides a convenient framework for enforcing
the interfacial conditions and constructing the dispersion relation. We seek axisymmetric normal modes of the form
\begin{equation}\label{eq:normal_modes_aw}
  \qty(u'_j,w'_j,p'_j)=\qty(\hat u_j(r),\hat w_j(r),\hat p_j(r))e^{\ii k z+s t},
  \qquad
  R_j(z,t)=\bar R_j+\eta_j e^{\ii k z+s t},
  \qquad j=1,2,3,
\end{equation}
where the prime quantities are disturbances from the base state.

\subsection{Radial-velocity formulation and the cylindrical Orr--Sommerfeld 
  equations}\label{sec:aw_os}
The incompressibility condition (\ref{eq:continuity}) gives
\begin{equation}\label{eq:aw_cont}
  \hat u_{j,r}+\frac{\hat u_j}{r}+\ii k\hat w_j=0,
\end{equation}
so that
\begin{equation}\label{eq:aw_w_from_u}
  \hat w_j(r)=-\frac{1}{\ii k}\qty(\hat u_{j,r}+\frac{\hat u_j}{r}).
\end{equation}
The linearised Stokes equations \eqref{eq:NS_r} and \eqref{eq:NS_z} in each layer are
\begin{equation}
  - \frac{\partial \hat p_j}{\partial r}
  + \frac{1}{m_j} \mathcal{L}_1 \hat u_j = 0,
  \label{eq:aw_radial_mom}
\end{equation}
and
\begin{equation}
  - \ii k \hat p_j
  + \frac{1}{m_j} \left( \mathcal{L}_1 + \frac{1}{r^2} \right) \hat w_j = 0,
  \label{eq:aw_axial_mom}
\end{equation}
where the differential operator $\mathcal{L}_1$ is defined by
\begin{equation}
  \mathcal{L}_1
  := \frac{\partial^2}{\partial r^2}g
  + \frac{1}{r}\frac{\partial}{\partial r}
  - \frac{1}{r^2}
  - k^2.
  \label{eq:aw_L1}
\end{equation}
Eliminating $\hat w_j$ using \eqref{eq:aw_w_from_u} and then eliminating $\hat p_j$ from 
\eqref{eq:aw_radial_mom}--\eqref{eq:aw_axial_mom} yields the fourth-order equation
\begin{equation}\label{eq:aw_os_eq}
  \mathcal{L}_1^2\hat u_j=0,
  \qquad \bar R_{j-1}<r<\bar R_j,
\end{equation}
with the convention $\bar R_0:=0$ and $\bar R_4:=a$ where needed. The general solution for (\ref{eq:aw_os_eq}) is
\begin{equation}\label{eq:aw_u_basis}
  \hat u_j(r)=A_j I_1(kr)+B_j K_1(kr)+C_j\,r I_0(kr)+D_j\,r K_0(kr),
  \qquad j=1,2,3,
\end{equation}
where $I_n$ and $K_n$ are modified Bessel functions. Substituting \eqref{eq:aw_u_basis} into \eqref{eq:aw_w_from_u} gives the axial velocity amplitudes. The pressure amplitudes follow from the radial momentum equation \eqref{eq:aw_radial_mom} as
\begin{equation}\label{eq:aw_p_from_u}
  \hat p_j(r)=\frac{2}{m_j} \qty(C_j I_0(kr)+D_j K_0(kr)).
\end{equation}

\subsection{Wall and interfacial conditions in terms of $\hat u_j$}\label{sec:aw_bcs}
At the wall $r=a$, the no-slip condition \eqref{eq:wall_bc} gives
\begin{equation}\label{eq:aw_wall_bc}
  \hat u_3(a)=0,
  \qquad
  \hat w_3(a)=0.
\end{equation}
At the inner interface $r=\bar R_1$, the tangential stress condition \eqref{eq:tan_free} yields
\begin{equation}\label{eq:aw_shearfree}
  \ii k\hat u_1(\bar R_1)+\hat w_{1,r}(\bar R_1)+\tau_1'(\bar R_1)\eta_1=0.
\end{equation}
The corresponding normal stress condition, (\ref{eq:norm_free}), is
\begin{equation}\label{eq:aw_N1}
  -\hat p_1(\bar R_1)+\frac{2}{m_1} \qty(\hat u_{1,r}(\bar R_1)-\ii k\tau_1(\bar R_1)\eta_1)
  -\frac{1}{\sigma_1 C}\qty(k^2-\frac{1}{\bar R_1^2})\eta_1=0.
\end{equation}
At $r=\bar R_2$ we impose continuity of radial velocity, (\ref{eq:vel_cont}),
\begin{equation}\label{eq:aw_ucont_2}
  \hat u_1(\bar R_2)-\hat u_2(\bar R_2)=0,
\end{equation}
continuity of axial velocity, (\ref{eq:vel_cont}),
\begin{equation}\label{eq:aw_wcont_2}
  \hat w_1(\bar R_2)-\hat w_2(\bar R_2)+\qty(\tau_1(\bar R_2)-\tau_2(\bar R_2))\eta_2=0,
\end{equation}
continuity of tangential stress, (\ref{eq:tangl_stress}),
\begin{equation}\label{eq:aw_tstress_2}
  m_2\qty(\ii k\hat u_1(\bar R_2)+\hat w_{1,r}(\bar R_2))
  -m_1\qty(\ii k\hat u_2(\bar R_2)+\hat w_{2,r}(\bar R_2))
  +\qty(m_2\tau_1'(\bar R_2)-m_1\tau_2'(\bar R_2))\eta_2=0,
\end{equation}
and continuity of normal stress, (\ref{eq:norm_stress}),
\begin{equation}\label{eq:aw_nstress_2}
  \qty[-\hat p_2+\frac{2}{m_2}\qty(\hat u_{2,r}-\ii k\tau_2\eta_2)]_{r=\bar R_2}
  -
  \qty[-\hat p_1+\frac{2}{m_1}\qty(\hat u_{1,r}-\ii k\tau_1\eta_2)]_{r=\bar R_2}
  -
  \frac{1}{\sigma_2 C}\qty(k^2-\frac{1}{\bar R_2^2})\eta_2=0.
\end{equation}

At $r=\bar R_3$ the corresponding conditions are
\begin{equation}\label{eq:aw_ucont_3}
  \hat u_2(\bar R_3)-\hat u_3(\bar R_3)=0,
\end{equation}
\begin{equation}\label{eq:aw_wcont_3}
  \hat w_2(\bar R_3)-\hat w_3(\bar R_3)+\qty(\tau_2(\bar R_3)-\tau_3(\bar R_3))\eta_3=0,
\end{equation}
\begin{equation}\label{eq:aw_tstress_3}
  \qty(\ii k\hat u_2(\bar R_3)+\hat w_{2,r}(\bar R_3))
  -m_2\qty(\ii k\hat u_3(\bar R_3)+\hat w_{3,r}(\bar R_3))
  +\qty(\tau_2'(\bar R_3)-m_2\tau_3'(\bar R_3))\eta_3=0,
\end{equation}
and
\begin{equation}\label{eq:aw_nstress_3}
  \qty[-\hat p_3+\frac{2}{m_3} \qty(\hat u_{3,r}-\ii k\tau_3\eta_3)]_{r=\bar R_3}
  -
  \qty[-\hat p_2+\frac{2}{m_2} \qty(\hat u_{2,r}-\ii k\tau_2\eta_3)]_{r=\bar R_3}
  -
  \frac{1}{C}\qty(k^2-\frac{1}{\bar R_3^2})\eta_3=0.
\end{equation}

The twelve non-kinematic conditions
\eqref{eq:aw_wall_bc}--\eqref{eq:aw_nstress_3} form a linear system for the coefficients
\[
  (A_1,B_1,C_1,D_1,\ldots,A_3,B_3,C_3,D_3)^T,
\]
in which the interface amplitudes $\eta_1,\eta_2,\eta_3$ appear as forcing terms.

\subsection{Kinematic closure and dispersion relation}\label{sec:aw_kin}

The coefficients in \eqref{eq:aw_u_basis} have been expressed in
§\ref{sec:aw_bcs} as linear functions of the interface amplitudes
$\eta_1,\eta_2,\eta_3$. Substituting these expressions into the
kinematic conditions
\begin{equation}
  s\eta_j-\hat u_j(\bar R_j)+\ii k \bar w_j \eta_j =0,
  \qquad j=1,2,3,
\end{equation}
yields a closed linear system for
$(\eta_1,\eta_2,\eta_3)$ of the form
\begin{equation}\label{eq:aw_reduced}
  \mathsf{M}(k,s)
  \begin{pmatrix}
    \eta_1\\
    \eta_2\\
    \eta_3
  \end{pmatrix}
  =\vect{0},
\end{equation}
where $\mathsf{M}(k,s)$ is a $3\times3$ matrix. Non-trivial solutions exist only if
\begin{equation}\label{eq:aw_dispersion}
  \det \mathsf{M}(k,s)=0,
\end{equation}
which defines the dispersion relation. Since the Stokes problem is
inertialess, the growth rate $s$ enters only through the kinematic
conditions, and \eqref{eq:aw_dispersion} is cubic in $s$ for each
fixed wavenumber $k$.

The results of the arbitrary wavenumber formulation presented here have been checked by 
comparison with a numerical solution of the linearised problem using a Chebyshev collocation 
method. Excellent agreement is obtained across the range of parameters considered.

\subsection{Influence of the middle-layer thickness $\bar{h}_2$}

The effect of the middle-layer thickness $\bar{h}_2$ on the linear stability is examined.
Throughout this section, the total thickness of the inner and middle layers is fixed such that
\begin{equation}
  \bar{h}_1 + \bar{h}_2 = 1,
\end{equation}
so that increasing $\bar{h}_2$ corresponds to a decrease in $\bar{h}_1$.
This parametrisation isolates the effect of redistributing fluid between the inner and middle layers while keeping their combined thickness fixed. For each value of $\bar{h}_2$ and each wavenumber $k$, the dispersion relation 
\eqref{eq:aw_dispersion} yields three eigenvalues $s_j(k)$, $j=1,2,3$, corresponding to distinct modes of disturbance. We define
\begin{equation}
  s_{\max}(k) := \max_{1 \leq j \leq 3} \Re\big(s_j(k)\big),
\end{equation}
and refer to the mode attaining this maximum as the dominant mode. The sign and magnitude of 
$s_{\max}(k)$ determine the linear stability characteristics of the flow. The identity of the dominant 
mode may vary with $k$, reflecting a transition between modes associated with different interfaces 
and hence a change in the underlying instability mechanism.

\begin{figure}
  \centering
  \includegraphics[width=0.65\textwidth]{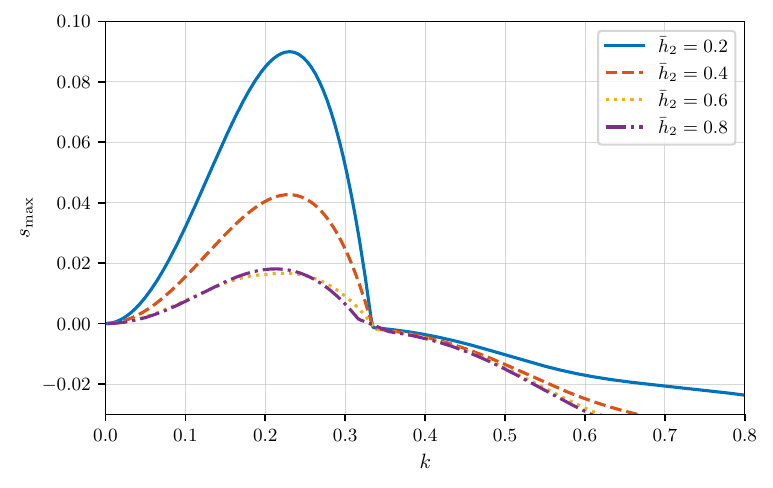}
  \caption{Largest growth rate $s_{\max}$ versus wavenumber $k$ for different values of 
    $\bar{h}_2$, with $m_1 = 50$ and $m_2 = 0.33$. The other parameters are $a=5$, $\sigma_1=\sigma_2=1$ and $C=0.3169$.}
 \label{fig:fig4}
\end{figure}

Figure~\ref{fig:fig4} shows $s_{\max}$ as a function of $k$ for several values of $\bar h_2$, with $m_1=50$ and $m_2=0.33$. In this parameter regime the unstable band is narrow and its location is nearly unchanged as fluid is redistributed between layers 1 and 2. Both the cut-off wavenumber and the most unstable wavenumber vary only weakly with $\bar h_2$; the principal effect of changing $\bar h_2$ is instead to alter the magnitude of the growth rate.

\begin{figure}
  \centering
  \includegraphics[width=0.65\textwidth]{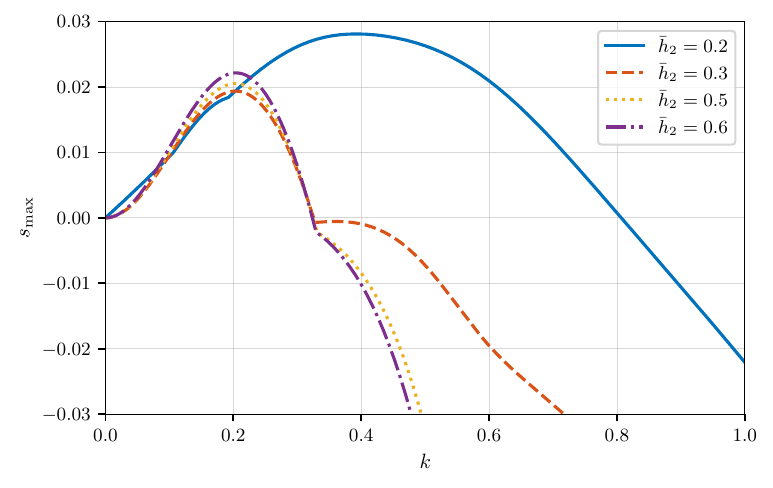}
  \caption{Largest growth rate $s_{\max}$ versus wavenumber $k$ for different values of 
    $\bar{h}_2$, with $m_1 = 0.5$ and $m_2 = 2$. The other parameters are the same as in figure~\ref{fig:fig4}.}
 \label{fig:fig5}
\end{figure}

To examine the role of viscosity contrast, we consider a second set of parameters with $\mu_1^* = 
2\,\mathrm{P}$, $\mu_2^* = 0.5\,\mathrm{P}$, and $\mu_3^* = 1\,\mathrm{P}$, corresponding to 
$m_1 = 0.5$ and $m_2 = 2$. The resulting growth rates are shown in figure~\ref{fig:fig5}, where the 
dependence on $\bar{h}_2$ is significantly more pronounced. For larger values of $\bar{h}_2$ (approximately $\bar{h}_2 \gtrsim 0.4$), the most unstable 
disturbances occur at small wavenumbers, with a peak near $k \approx 0.2$, indicating predominantly long-wave behaviour. As $\bar{h}_2$ decreases, the most unstable wavenumber shifts to larger values, and shorter-wavelength disturbances become increasingly important. At the same time, the maximum growth rate varies non-monotonically with $\bar{h}_2$, indicating a change in the underlying instability mechanism. The transition occurs near $\bar h_2 \approx 0.35$. For smaller values of $\bar h_2$, the
instability extends over a broad range of wavenumbers and is dominated by
shorter-wavelength disturbances, whereas for larger values of $\bar h_2$ it
becomes increasingly confined to the long-wave regime.

\begin{figure}
  \centering
  \includegraphics[width=0.65\textwidth]{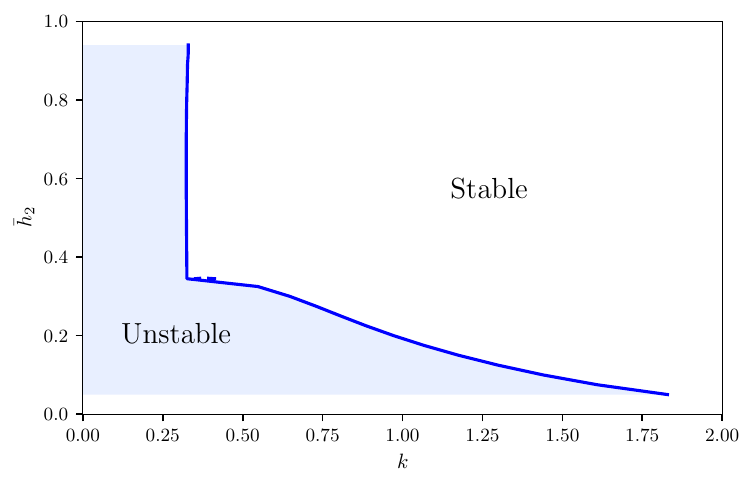}
  \caption{Stability diagram in the $(\bar h_2,k)$-plane showing the cut-off
    wavenumber separating stable and unstable regions. The flow is unstable
    for wavenumbers below the cut-off curve and stable above it.
    Parameters are $a=5$, $m_1=0.5$, $m_2=2$, and
    $\sigma_1=\sigma_2=1$.}
 \label{fig:fig6}
\end{figure}

To characterise the stability boundaries, we compute the cut-off wavenumber
$k_0$ defined by $s_{\max}(k_0)=0$. The resulting stability diagram in the
$(\bar h_2,k)$-plane is shown in figure~\ref{fig:fig6}. The parameters are
$a=5$, $\bar h_1=0.8$, $\bar h_3=1$, $m_1=0.5$, $m_2=2$, $\sigma_1=\sigma_2=1$,
with $\bar h_2\in[0.01,0.8]$.
The system is unstable for sufficiently small wavenumbers, $0 < k < k_{0l}$ with $k_{0l} \approx 
0.3$, for all values of $\bar{h}_2$, indicating a robust long-wave instability. For larger wavenumbers, 
the stability boundary depends on $\bar{h}_2$. In particular, for smaller values of $\bar{h}_2$ 
(approximately $\bar{h}_2 \lesssim 0.35$), the cut-off wavenumber increases, leading to a broader 
unstable band. As $\bar{h}_2$ increases, this band contracts and the instability becomes confined 
primarily to the long-wave regime.

The robustness of this long-wave band, and in particular the near-constancy of
its lower cut-off $k_{0l}$, has a simple capillary origin. Under the constraint
$\bar h_1+\bar h_2=1$ (with $\bar h_3=1$ and $a=5$), redistributing fluid
between layers~1 and~2 leaves the inner and outer interface radii fixed,
$\bar R_1=a-(\bar h_1+\bar h_2+\bar h_3)=3$ and $\bar R_3=a-\bar h_3=4$, and
moves only the middle interface $\bar R_2=4-\bar h_2$. For an air--liquid
interface of radius $\bar R_1$, azimuthal (Rayleigh--Plateau) curvature
destabilises disturbances with $k<1/\bar R_1$, the neutral wavenumber
$k=1/\bar R_1$ marking the balance between the destabilising azimuthal
curvature $1/\bar R_1^2$ and the stabilising axial curvature $k^2$
\citep{Goren1962,Hammond1983}. The computed lower cut-off lies just below this
value throughout, $k_{0l}\approx0.323$--$0.330$ against $1/\bar R_1\approx0.333$,
varying by only a few percent; the small offset and weak, non-monotonic
dependence on $\bar h_2$ reflects the residual coupling to the middle interface,
whose radius $\bar R_2$ is the only one that changes under the constraint. The
persistent long-wave band is therefore the capillary instability of the
air--core interface, essentially pinned by the fixed inner radius $\bar R_1$,
while the structure at larger $k$, including the narrow secondary band
near $\bar h_2\approx0.345$, which marks the transition to the broad-band
regime at smaller $\bar h_2$, is consistent with the increasing influence of
interfacial coupling.

\subsection{Influence of the viscosity ratio $m_1$}

We examine the influence of the viscosity ratio $ m_1 = \mu_3^*/\mu_1^*$,
which measures the relative viscosity of the outer layer (layer 3) to that of the inner layer (layer 1). 
The chosen dimensional parameters are $a^* = 0.5\,\mathrm{cm}$, $\bar{h}_1^* = 
0.08\,\mathrm{cm}$, $\bar{h}_2^* = 0.02\,\mathrm{cm}$, $\bar{h}_3^* = 0.1\,\mathrm{cm}$, 
$\sigma_1^* = \sigma_2^* = \sigma_3^* = 21.5\,\mathrm{dyn\,cm^{-1}}$, $\rho^* = 
0.97\,\mathrm{g\,cm^{-3}}$, $\mu_3^* = 4\,\mathrm{P}$, and $\mu_2^* = 2\,\mathrm{P}$. The viscosity ratio $m_1$ is varied through $\mu_1^*$, with all other parameters fixed. The corresponding dimensionless parameters are $a = 5$, $\bar{h}_1 = 0.8$, $\bar{h}_2 = 0.2$, $\bar{h}_3 = 1$, $C = 0.3169$, and $m_2 = 2$.

\begin{figure}
  \centering
  \includegraphics[width=0.65\textwidth]{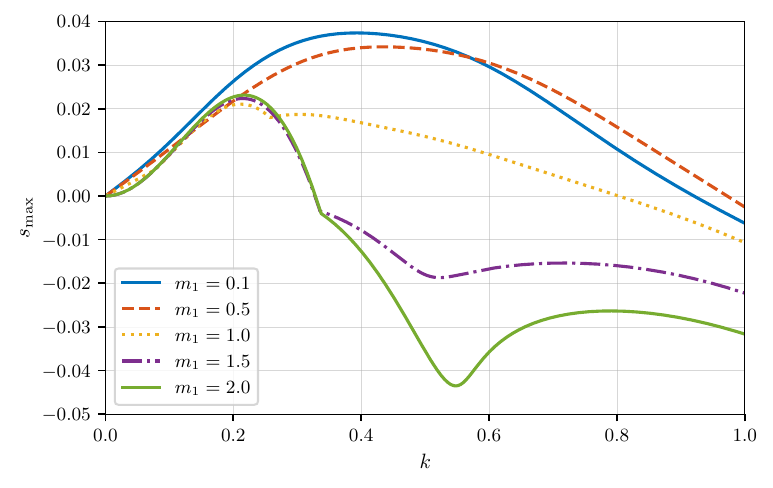}
  \caption{Largest growth rate $s_{\max}$ versus wavenumber $k$ for different values of 
    $m_1$, with $a = 5$, $\bar{h}_1 = 0.8$, $\bar{h}_2 = 0.2$, 
    $\bar{h}_3 = 1$, $C = 0.3169$, and $m_2 = 2$.}
 \label{fig:fig7}
\end{figure}

Figure~\ref{fig:fig7} shows $s_{\max}(k)$ for several values of $m_1$. The effect of $m_1$ depends strongly on whether its value is less than or greater than unity. For $m_1 \lesssim 1$, corresponding to an inner layer 
more viscous than the outer layer, decreasing $m_1$ leads to a pronounced increase in the maximum growth rate and a widening of the unstable wavenumber range. In this regime, the instability extends to larger $k$, indicating increased sensitivity to shorter-wavelength disturbances. For $m_1 \gtrsim 1$, corresponding to a less viscous inner layer, the growth rate curves depend only 
weakly on $m_1$. Both the maximum growth rate and the most unstable wavenumber vary only 
modestly as $m_1$ increases.These results indicate that the instability is considerably more sensitive to changes in $m_1$ when the inner layer
is more viscous than the outer layer ($m_1<1$).

\begin{figure}
  \centering
  \includegraphics[width=0.65\textwidth]{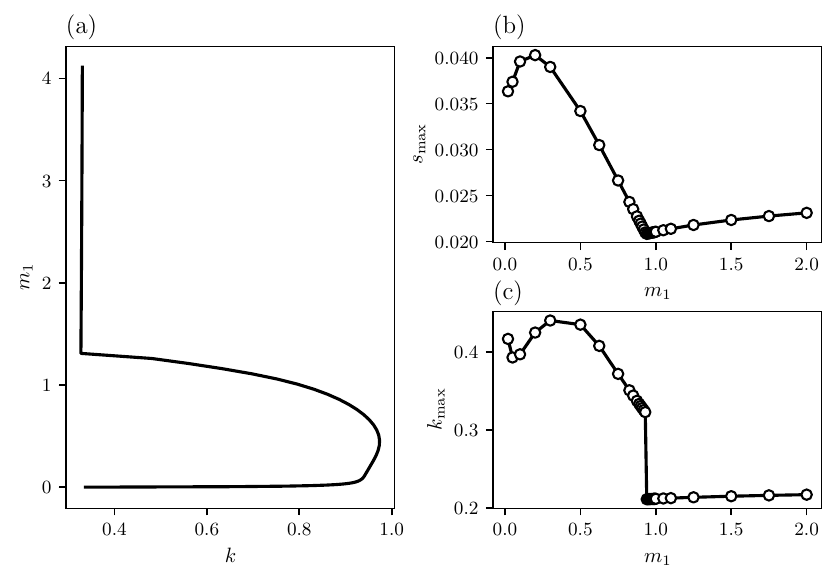}
  \caption{(a) Stability regions in the $(m_1,k)$-plane. (b) Maximum growth rate 
    $s_{\max}$ as a function of $m_1$. (c) Most unstable wavenumber $k_{\max}$ as a function 
    of $m_1$.}
 \label{fig:fig8}
\end{figure}

Figure~\ref{fig:fig8} gives a summary of the influence of $m_1$ on both the unstable wavenumber range and the dominant growth rate. Panel (a) shows that the system is unstable for sufficiently small wavenumbers, $0 < k < k_{0l}$ with 
$k_{0l} \approx 0.3$, for all values of $m_1$, indicating a robust long-wave instability. As $m_1$ decreases from values near unity, the unstable wavenumber interval expands, reaching its widest extent near $m_1 \approx 0.5$, before shrinking again for smaller $m_1$. Thus the dependence of the cut-off wavenumber on $m_1$ is non-monotonic.

Panels (b) and (c) show that $s_{\max}$ varies smoothly with $m_1$, attaining its largest values for 
intermediate $m_1 \approx 0.2$--$0.3$, and decreasing toward a minimum near $m_1 \approx 1$. In contrast, the most unstable wavenumber, $k_{\max}$, exhibits a sharp transition near $m_1 \approx 1$. Specifically, $k_{\max}$ 
decreases gradually as $m_1$ increases from small values, but undergoes a rapid shift from $k_{\max} \approx 0.32$ to $k_{\max} \approx 0.21$ as $m_1$ approaches unity. The abrupt transition in $k_{\max}$ is a consequence of the coexistence of two competing local maxima of $s_{\max}(k)$. For $m_1$ slightly below unity, the growth-rate curve exhibits two peaks of comparable magnitude at different wavenumbers. As $m_1$ varies, the global maximum switches between these peaks, leading to a discontinuous change in $k_{\max}$ while $s_{\max}$ remains continuous.

Overall, varying the viscosity ratio $m_1$ modifies both the strength and the
character of the instability. In addition to changing the maximum growth rate
and the extent of the unstable wavenumber range, it can also produce an abrupt
transition in the dominant unstable wavelength through competition between
distinct instability modes. These results demonstrate that the viscous contrast
between the inner and outer liquid layers plays a central role in determining
both the dominant instability mechanism and the resulting linear stability
characteristics.

\subsection{Influence of the surface tension ratio $\sigma_1$}

We examine the influence of the surface tension ratio $\sigma_1 =\sigma_3^*/\sigma_1^*$, which measures the relative strength of surface tension at the inner interface compared to that at the outer interface. The analysis is performed for three representative viscosity pairs $(m_1,m_2)$ 
spanning the principal regimes identified in the previous subsection. The dimensional surface tensions are chosen such that $\sigma_2^* = \sigma_3^* = 
21.5\,\mathrm{dyn\,cm^{-1}}$, while $\sigma_1^*$ is varied over a physically relevant range 
corresponding to $\sigma_1 \in [0.05,\,10]$. The remaining parameters are $a = 5$, $h_1 = 0.8$, 
$h_2 = 0.2$, and $h_3= 0.1$. Three different pairs of viscosity ratios are considered here: 
$(m_1,m_2) = (2,2)$; $(0.5,2)$; and $(0.5,0.5)$.

\begin{figure}
  \centering
  \includegraphics[width=0.85\textwidth]{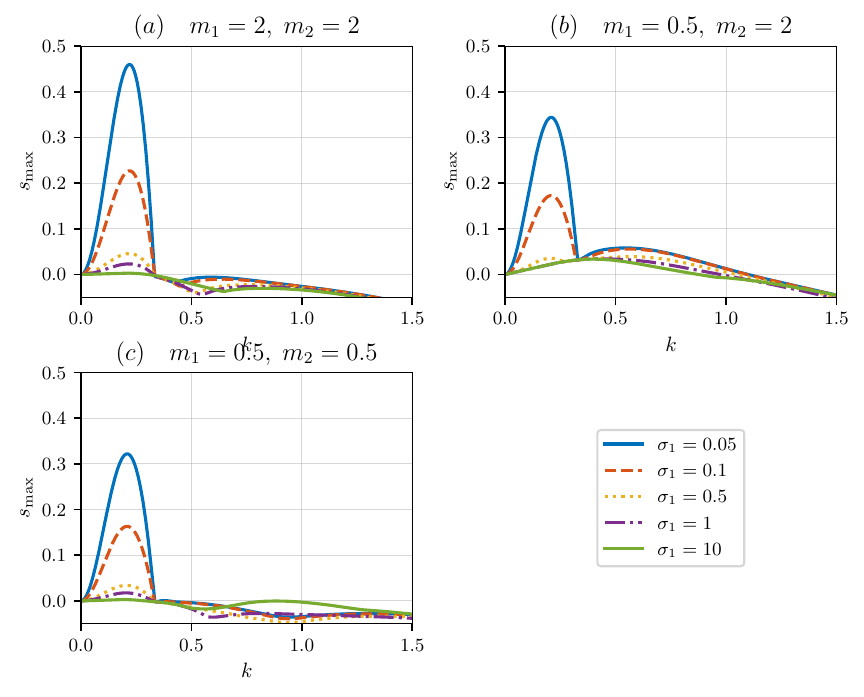}
  \caption{Largest growth rate $s_{\max}$ versus wavenumber $k$ for different values of 
    $\sigma_1$ and viscosity ratios: (a) $(m_1,m_2)=(2,2)$, 
    (b) $(m_1,m_2)=(0.5,2)$, and (c) $(m_1,m_2)=(0.5,0.5)$.}
 \label{fig:fig9}
\end{figure}

For all three viscosity configurations shown in figure~\ref{fig:fig9}, increasing $\sigma_1$ reduces the
growth rate across much of the unstable wavenumber range, but the response
depends strongly on the viscosity ratios.
 In panel (a), where $(m_1,m_2)=(2,2)$, the instability is concentrated at small wavenumbers, with the largest growth rates occurring near the long-wave end of the spectrum. In panel (b), where $(m_1,m_2)=(0.5,2)$, the unstable interval is much broader, with marginal stability occurring near $k\simeq 1$ for the smaller values of $\sigma_1$. Thus, in this viscosity regime, intermediate-wavelength disturbances remain important even though increasing $\sigma_1$ reduces the growth rate. The case $(m_1,m_2)=(0.5,0.5)$, shown in panel (c), exhibits a different
structure. For sufficiently small $\sigma_1$, the growth-rate curve contains
both the long-wave unstable band and an additional unstable window at moderate
wavenumbers. Increasing $\sigma_1$ suppresses this secondary window, so that
the instability becomes confined primarily to the long-wave regime. Thus the
surface-tension ratio does not merely change the magnitude of the growth
rate; it can also change the number and location of unstable bands.

\begin{figure}
  \centering
  \includegraphics[width=0.65\textwidth]{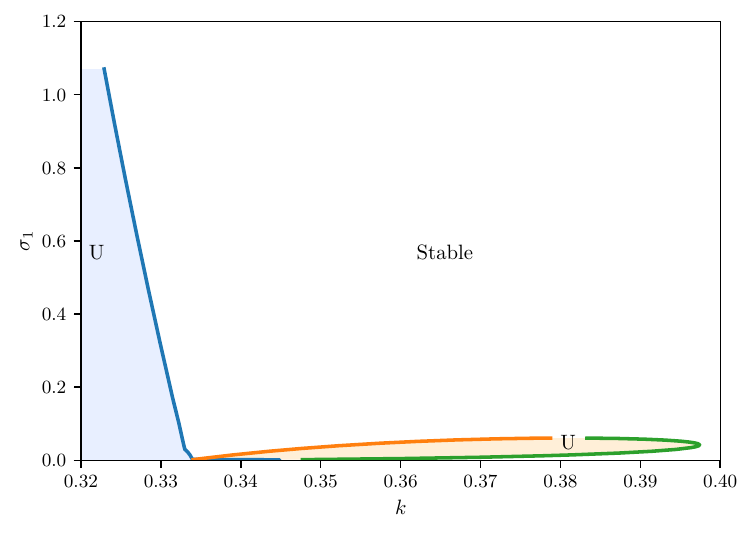}
  \caption{Stability regions in the $(\sigma_1,k)$-plane for $(m_1,m_2) = (0.5,0.5)$.}
 \label{fig:fig10}
\end{figure}

To quantify the stability boundaries, we compute the cut-off wavenumber $k_0$ defined by $s_{\max}(k_0)=0$. The dependence of $k_0$ on $\sigma_1$ is shown in figure~\ref{fig:fig10} for the representative case $(m_1,m_2) = (0.5,0.5)$.
The system is unstable for sufficiently small wavenumbers, $0 < k < k_{0l}$ with $k_{0l} \approx 
0.33$, for all values of $\sigma_1$, indicating a robust long-wave instability. For sufficiently small 
$\sigma_1$, a second, disconnected unstable band appears at moderate wavenumbers, 
giving rise to two distinct 
instability regions. This secondary instability is not observed for the viscosity configurations in 
figure~\ref{fig:fig9}(a) and (b), and appears to be specific to $(m_1,m_2) = (0.5,0.5)$. As $\sigma_1$ increases, this secondary unstable band disappears and the instability becomes restricted to the long-wave regime.

Overall, increasing the surface tension at the inner interface suppresses
short- and intermediate-wavelength disturbances and eventually eliminates the
secondary instability altogether, leaving only the robust long-wave
instability. 

\subsection{Summary of arbitrary-wavenumber results}

The arbitrary-wavenumber analysis reveals the distinct roles played by layer
thickness, viscosity contrast, and surface tension in determining the linear
stability characteristics of the three-layer flow.

Variations in the middle-layer thickness $\bar h_2$ primarily affect the
coupling between the two liquid interfaces. Increasing $\bar h_2$ suppresses
the broad-band instability at moderate and large wavenumbers, leaving the
persistent long-wave instability as the dominant mode.

The viscosity ratio $m_1$ controls the strength of viscous coupling at the inner interface and 
produces a strongly asymmetric response. For $m_1 \lesssim 1$, decreasing $m_1$ enhances both 
the growth rate and the range of unstable wavenumbers, while for $m_1 \gtrsim 1$ the dependence 
is comparatively weak. Near $m_1 \approx 1$, competing local maxima in the growth rate lead to a 
transition in the most unstable wavenumber, associated with a change in the dominant instability 
mode.

Overall, variations in the surface tension ratio $\sigma_1$ influence both the
strength and the structure of the instability. Since
$\sigma_1=\sigma_3^*/\sigma_1^*$, increasing $\sigma_1$ corresponds to
decreasing the surface tension at the air--core interface. Over the parameter
range considered, this leads to a reduction in the maximum growth rate and progressively suppresses the secondary finite-wavenumber unstable band, leaving only the persistent long-wave instability for sufficiently large $\sigma_1$. Thus, increasing $\sigma_1$ both weakens the instability and contracts the unstable spectrum towards the long-wave regime.

Taken together, these results indicate that the instability arises from a balance between viscous 
stratification, interfacial coupling, and capillary effects. Despite the complexity of the 
finite-wavenumber structure, including mode switching and secondary unstable bands, a robust 
long-wave instability persists across all parameter regimes considered.

The persistence of the long-wave instability throughout the parameter space
suggests that a reduced asymptotic description in the long-wavelength limit
should capture the principal instability responsible for the onset of
nonlinear evolution, while the arbitrary-wavenumber analysis identifies
additional finite-wavenumber features that lie beyond the scope of the
asymptotic theory.

\section{Long-wave approximation}

The arbitrary-wavenumber analysis in §3 shows that the dominant
instability is typically associated with small axial wavenumbers
across a broad range of parameter values. This motivates the
development of a reduced asymptotic model in the long-wave limit.
Using the long-wave parameter $\epsilon \ll 1$ introduced in §2.4,
we now derive an evolution system describing the weakly varying
interfacial dynamics.

We seek asymptotic expansions of the dependent variables in powers of $\epsilon$:
\begin{align*}
  u_j &= u_j^{(0)} + \epsilon u_j^{(1)} + \epsilon^2 u_j^{(2)} + \cdots,\\
  w_j &= w_j^{(0)} + \epsilon w_j^{(1)} + \epsilon^2 w_j^{(2)} + \cdots,\\
  p_j &= p_j^{(0)} + \epsilon p_j^{(1)} + \epsilon^2 p_j^{(2)} + \cdots.
\end{align*}

At leading order, the pressure is independent of the radial coordinate in each layer, while the axial 
velocity is determined by a balance between viscous stresses and the gravitational body force. This 
yields lubrication-type flow profiles in each layer. Substituting these expansions into the governing equations and retaining the
leading-order terms, we obtain, for $j=1,2,3$,
\begin{equation}
  p_{j,r}^{(0)}=0,\qquad
  \frac{1}{r}\partial_r\!\left(r w_{j,r}^{(0)}\right)=-m_j,\qquad
  \frac{1}{r}\partial_r\!\left(r u_j^{(0)}\right)
  +w_{j,z}^{(0)}=0.
  \label{eq:lo}
\end{equation}
Here the constant right-hand side of \eqref{eq:lo}$b$ arises from gravity after
nondimensionalisation.

The corresponding leading-order boundary and interfacial conditions are as follows. At the tube wall 
$r=a$, the no-slip and no-penetration conditions are
\begin{equation}
  u_3^{(0)} = 0, \qquad w_3^{(0)} = 0.
  \label{eq:lo_wall}
\end{equation}
At the outer interface $r=R_3(z,t)$, we impose continuity of tangential stress, the normal-stress 
jump, the kinematic condition, and continuity of velocity:
\begin{align}
  m_2 w_{3,r}^{(0)} &= w_{2,r}^{(0)}, \qquad
                      p_2^{(0)} - p_3^{(0)} = \frac{1}{C}\left(\frac{1}{R_3} - \epsilon^2 R_{3,zz}\right),
                      \label{eq:lo_R3_stress}\\
  u_2^{(0)} &= R_{3,t} + w_2^{(0)} R_{3,z}, \qquad
              u_2^{(0)} = u_3^{(0)}, \qquad w_2^{(0)} = w_3^{(0)}.
              \label{eq:lo_R3_kin}
\end{align}
Similar conditions are applied at the interface $r=R_2(z,t)$:
\begin{align}
  m_1 w_{2,r}^{(0)} &= m_2 w_{1,r}^{(0)}, \qquad
                      p_1^{(0)} - p_2^{(0)} = \frac{1}{\sigma_2 C}
                      \left(\frac{1}{R_2} - \epsilon^2 R_{2,zz}\right),
                      \label{eq:lo_R2_stress}\\
  u_1^{(0)} &= R_{2,t} + w_1^{(0)} R_{2,z}, \qquad
              u_1^{(0)} = u_2^{(0)}, \qquad w_1^{(0)} = w_2^{(0)}.
              \label{eq:lo_R2_kin}
\end{align}
Finally, at the free surface $r=R_1(z,t)$, the conditions are
\begin{equation}
  w_{1,r}^{(0)} = 0, \qquad
  -p_1^{(0)} + p_{\mathrm{atm}} = \frac{1}{\sigma_1 C}
  \left(\frac{1}{R_1} - \epsilon^2 R_{1,zz}\right), \qquad
  u_1^{(0)} = R_{1,t} + w_1^{(0)} R_{1,z}.
  \label{eq:lo_R1}
\end{equation}
As in \citet{Ogrosky2021}, we retain the formally higher-order curvature terms $\epsilon^2 R_{i,zz}$ 
in the normal-stress conditions. These terms are required to capture the stabilising effect of surface 
tension at shorter wavelengths and, in particular, to recover a finite cut-off wavenumber in the linear theory. The leading-order axial velocities are obtained by integrating the axial momentum equations
\eqref{eq:lo}$b$ subject to the wall, interfacial, and continuity conditions.
The corresponding radial velocities are then determined from the
continuity equations \eqref{eq:lo}$c$
together with the kinematic and no-penetration conditions.
The resulting leading-order velocity fields are
\begin{multline}
  u_1^{(0)}
  =
  \frac{R_1R_{1,z}}{4r}
  \Bigg[
  -a^2+m_1r^2+(m_2-m_1)R_2^2-(m_2-1)R_3^2
  \\
  +2r^2
  \left(
    \ln a
    -m_1\ln r
    +(m_1-m_2)\ln R_2
    +(m_2-1)\ln R_3
  \right)
  \Bigg]
  \\
  +
  \frac{(m_1-m_2)(r^2-R_2^2)(R_1^2-R_2^2)}
  {4rR_2}
  R_{2,z}
  \\
  +
  \frac{(m_2-1)(r^2-R_3^2)(R_1^2-R_3^2)}
  {4rR_3}
  R_{3,z},
  \label{eq:u1}
\end{multline}
\begin{multline}
  w_1^{(0)} = \frac{1}{4}\left[a^2 + (m_2-1)R_3^2 + (m_1-m_2)R_2^2 - m_1 r^2\right] \\
  + \frac{m_1 R_1^2}{2}\ln\frac{r}{R_2}
  + \frac{m_2 R_1^2}{2}\ln\frac{R_2}{R_3}
  + \frac{R_1^2}{2}\ln\frac{R_3}{a},
  \label{eq:w1}
\end{multline}
\begin{multline}
  u_2^{(0)}
  =
  \frac{R_1R_{1,z}}{4r}
  \Bigg[
  -a^2+m_2r^2-(m_2-1)R_3^2
  \\
  +2r^2
  \left(
    \ln a
    -m_2\ln r
    +(m_2-1)\ln R_3
  \right)
  \Bigg]
  \\
  +
  \frac{(m_2-1)(r^2-R_3^2)(R_1^2-R_3^2)}
  {4rR_3}
  R_{3,z}.
  \label{eq:u2}
\end{multline}
\begin{equation}
  w_2^{(0)} = \frac{1}{4}\left[a^2 + (m_2-1)R_3^2 - m_2 r^2\right]
  + \frac{m_2 R_1^2}{2}\ln\frac{r}{R_3}
  + \frac{R_1^2}{2}\ln\frac{R_3}{a},
  \label{eq:w2}
\end{equation}
\begin{equation}
  u_3^{(0)} = -\frac{1}{4r}\left(a^2 - r^2 + 2r^2\ln\frac{r}{a}\right) R_1 R_{1,z},
  \label{eq:u3}
\end{equation}
\begin{equation}
  w_3^{(0)} = \frac{1}{4}\left(a^2 - r^2\right) + \frac{R_1^2}{2}\ln\frac{r}{a}.
  \label{eq:w3}
\end{equation}
The axial velocity fields correspond to generalised Poiseuille
profiles modified by the interfacial geometry and viscous coupling
between the liquid layers. In the special case $m_1=m_2=1$, the velocity fields reduce to the corresponding single-viscosity expressions.

The leading-order pressures follow directly from the normal-stress balances 
\eqref{eq:lo_R3_stress}$b$, \eqref{eq:lo_R2_stress}$b$, and \eqref{eq:lo_R1}$b$.
\begin{align}
  p_1^{(0)} &= p_{\mathrm{atm}} - \frac{1}{\sigma_1 C}
              \left(\frac{1}{R_1} - \epsilon^2 R_{1,zz}\right),
              \label{eq:p1}\\
  p_2^{(0)} &= p_1^{(0)} - \frac{1}{\sigma_2 C}
              \left(\frac{1}{R_2} - \epsilon^2 R_{2,zz}\right),
              \label{eq:p2}\\
  p_3^{(0)} &= p_2^{(0)} - \frac{1}{C}
              \left(\frac{1}{R_3} - \epsilon^2 R_{3,zz}\right).
              \label{eq:p3}
\end{align}
We now consider the $O(\epsilon)$ problem, which provides the corrections required to determine the evolution of the interfaces. At this order, the velocity and pressure fields satisfy, for $j=1,2,3$,
\begin{equation}
  m_j p_{1,r}^{(1)} = \frac{1}{r}\partial_r\!\left(u_{j,r}^{(0)}\right) - \frac{u_j^{(0)}}{r^2},
  \quad
  \frac{1}{r}\partial_r\!\left(w_{j,r}^{(1)}\right) = m_j p_{j,z}^{(0)},
  \quad
  \frac{1}{r}\partial_r\!\left(u_j^{(1)}\right) + w_{j,z}^{(1)} = 0.
  \label{eq:o1}
\end{equation}
The corresponding boundary conditions are obtained by expanding the stress and kinematic conditions to $O(\epsilon)$. At the wall $r=a$,
\begin{equation}
  u_3^{(1)} = 0, \qquad w_3^{(1)} = 0.
  \label{eq:o1_wall}
\end{equation}
At $r=R_3(z,t)$, the first-order conditions are
\begin{equation}
  m_2 w_{3,r}^{(1)} = w_{2,r}^{(1)}, \qquad
                      p_2^{(1)} - p_3^{(1)} = \frac{2}{m_2}\left(u_{2,r}^{(0)} - m_2 u_{3,r}^{(0)}\right),
                      \label{eq:o1_R3_stress}
\end{equation}
and
\begin{equation}
  u_2^{(1)} = R_{3,t} + w_2^{(1)} R_{3,z}, \qquad
              u_2^{(1)} = u_3^{(1)}, \qquad w_2^{(1)} = w_3^{(1)},
              \label{eq:o1_R3_kin}
\end{equation}
while at $r=R_2(z,t)$, they are
\begin{equation}
  m_1 w_{2,r}^{(1)} = m_2\left(w_{1,r}^{(1)} + u_{1,z}^{(0)}\right), \qquad
                      p_1^{(1)} - p_2^{(1)} = \frac{2}{m_1 m_2}
                      \left(m_2 u_{1,r}^{(0)} - m_1 u_{2,r}^{(0)}\right),
                      \label{eq:o1_R2_stress}
\end{equation}
and
\begin{equation}
u_1^{(1)} = R_{2,t} + w_1^{(1)} R_{2,z}, \qquad
              u_1^{(1)} = u_2^{(1)}, \qquad w_1^{(1)} = w_2^{(1)}.
              \label{eq:o1_R2_kin}
\end{equation}
At the free surface, $r=R_1(z,t)$, the conditions are
\begin{equation}
  w_{1,r}^{(1)} = 0, \qquad
  p_1^{(1)} = \frac{2}{m_1}u_{1,r}^{(0)}, \qquad
  u_1^{(1)} = R_{1,t} + w_1^{(1)} R_{1,z}.
  \label{eq:o1_R1}
\end{equation}

The first-order corrections to the axial velocity are obtained by integrating 
\eqref{eq:o1}$b$ subject to the corresponding boundary and interfacial 
conditions, and are given by
\begin{multline}
  w_1^{(1)} = \frac{1}{4}\,p_{1,z}^{(0)}\left[m_1\left(r^2 - R_2^2
      - 2R_1^2\ln\frac{r}{R_2}\right)
    + 2\left(R_2^2 - R_1^2\right)\left(m_2\ln\frac{R_2}{R_3}
      - \ln\frac{a}{R_3}\right)\right] \\
  + \frac{1}{4}\,p_{2,z}^{(0)}\left[m_2\left(R_2^2 - R_3^2
      + 2R_2^2\ln\frac{R_3}{R_2}\right)
    + 2\left(R_2^2 - R_3^2\right)\ln\frac{a}{R_3}\right] \\
  + \frac{1}{4}\,p_{3,z}^{(0)}\left[R_3^2 - a^2 + 2R_3^2\ln\frac{a}{R_3}\right],
  \label{eq:w1_1}
\end{multline}
\begin{multline}
  w_2^{(1)} = \frac{1}{4}\,p_{1,z}^{(0)}\left[2\left(R_2^2 - R_1^2\right)
    \left(m_2\ln\frac{r}{R_3} - \ln\frac{a}{R_3}\right)\right] \\
  + \frac{1}{4}\,p_{2,z}^{(0)}\left[m_2\left(r^2 - R_3^2
      - 2R_2^2\ln\frac{r}{R_3}\right)
    + 2\left(R_2^2 - R_3^2\right)\ln\frac{a}{R_3}\right] \\
  + \frac{1}{4}\,p_{3,z}^{(0)}\left[R_3^2 - a^2 + 2R_3^2\ln\frac{a}{R_3}\right],
  \label{eq:w2_1}
\end{multline}
\begin{equation}
  w_3^{(1)} = \frac{1}{2}\,p_{1,z}^{(0)}\left(R_2^2 - R_1^2\right)\ln\frac{r}{a}
  +\frac{1}{2}\,p_{2,z}^{(0)}\left(R_3^2 - R_2^2\right)\ln\frac{r}{a}
  + \frac{1}{4}\,p_{3,z}^{(0)}\left[r^2 - a^2 - 2R_3^2\ln\frac{r}{a}\right].
  \label{eq:w3_1}
\end{equation}
The corresponding radial velocities follow from the continuity equations
and are required in establishing the kinematic flux relations, although
their explicit forms are not needed in the final evolution equations.

\subsection{Evolution equations}

The kinematic conditions are expressed in terms of volumetric fluxes in each liquid layer. We define
\begin{equation}
  Q_j(z,t) = \int_{R_j(z,t)}^{R_{j+1}(z,t)} r\,w_j(r,z,t)\,dr,
  \qquad j=1,2,3,\qquad R_4 = a,
  \label{eq:Qj_def}
\end{equation}
where $w_j$ denotes the axial velocity in layer $j$. Using incompressibility together with the 
kinematic boundary conditions yields
\begin{equation}
  \frac{1}{2}(R_3^2)_t = (Q_3)_z,\qquad
  \frac{1}{2}(R_2^2 - R_3^2)_t = (Q_2)_z,\qquad
  \frac{1}{2}(R_1^2 - R_2^2)_t = (Q_1)_z.
  \label{eq:kin_flux}
\end{equation}
It is convenient to introduce the cumulative fluxes
\begin{equation}
  Q^{(3)} := Q_3,\qquad
  Q^{(2)} := Q_2 + Q_3,\qquad
  Q^{(1)} := Q_1 + Q_2 + Q_3,
  \label{eq:Qcum}
\end{equation}
so that
\begin{equation}
  R_{j,t} = \frac{1}{R_j}\partial_z Q^{(j)}, \qquad j=1,2,3.
  \label{eq:Rj_flux_compact}
\end{equation}
To determine these fluxes, we substitute the asymptotic expansion
\begin{equation}
  w_j = w_j^{(0)} + \epsilon w_j^{(1)}, \qquad j=1,2,3,
\end{equation}
into \eqref{eq:Qj_def}. Following \citet{Ogrosky2021}, we rescale the axial coordinate so that $\epsilon$ does not appear explicitly in the resulting evolution equations. After straightforward but lengthy algebra, each cumulative flux may be written in the form
\begin{equation}
  Q^{(j)} = \sum_{k=1}^{3} A_{jk}(R_1,R_2,R_3)\,p_{k,z}^{(0)}
  + G_j(R_1,R_2,R_3), \qquad j=1,2,3,
  \label{eq:Q_flux_form}
\end{equation}
where $A_{jk}$ and $G_j$ depend on the interface positions and system parameters. The 
corresponding axial pressure gradients are obtained by
differentiating \eqref{eq:p1}--\eqref{eq:p3} with respect to $z$. Substituting \eqref{eq:Q_flux_form} 
into \eqref{eq:Rj_flux_compact} yields the closed system
\begin{equation}
  R_{j,t} = \frac{1}{R_j}\partial_z
  \left(
    \sum_{k=1}^{3} A_{jk} p_{k,z}^{(0)} + G_j
  \right), \qquad j=1,2,3.
  \label{eq:Rj_flux_final}
\end{equation}
Equations \eqref{eq:Rj_flux_final} constitute the long-wave evolution equations governing the three 
interfaces. Each equation has the form of a conservation law, with fluxes determined by
capillary pressure gradients, viscous coupling between the layers, and gravitational drainage.

The coefficient functions satisfy several consistency checks.
In particular, when \(R_2=R_1\), the inner layer has zero thickness so that
\(Q_1=0\), implying \(Q^{(1)}=Q^{(2)}\). The coefficient functions in
Appendix~B satisfy this identity through
\[
  \left. A_{1k}\right|_{R_2=R_1}
  =
  \left. A_{2k}\right|_{R_2=R_1},
  \qquad k=1,2,3,
\]
together with
\[
  \left. G_1\right|_{R_2=R_1}
  =
  \left. G_2\right|_{R_2=R_1}.
\]
In addition, the uniform base state \(R_j=\bar R_j\) recovers the steady
concentric drainage solution.

\subsection{Linear stability of the long-wave model}\label{sec:lw_linear}

We examine the linear stability of the reduced evolution equations derived in the previous section.
This analysis complements the arbitrary-wavenumber formulation of \S3.
The full Stokes formulation yields a dispersion relation containing modified
Bessel functions of the axial wavenumber, whereas the long-wave model gives its
small-wavenumber approximation. Specifically, in the limit $k\to0$, the
arbitrary-wavenumber dispersion relation reduces to that of the long-wave model,
provided the axial curvature contribution is retained.

\subsubsection{Linearisation}

We consider perturbations about the uniform base state
\begin{equation}
  R_j(z,t) = \bar{R}_j, \qquad j=1,2,3,
\end{equation}
where $\bar{R}_1 < \bar{R}_2 < \bar{R}_3 < a$ are constants. Normal-mode perturbations of the form are introduced
\begin{equation}
  R_j(z,t) = \bar{R}_j + \hat{R}_j e^{ikz + s t}, \qquad j=1,2,3,
  \label{eq:lw_modes}
\end{equation}
where, as defined earlier, $k$ is the axial wavenumber and $s$ is the complex growth rate. Substituting \eqref{eq:lw_modes} into the evolution equations
\begin{equation}
  R_{j,t} = \frac{1}{R_j}\partial_z Q^{(j)}, \qquad j=1,2,3,
\end{equation}
and linearising about the base state yields
\begin{equation}
  s \hat{R}_j = \frac{ik}{\bar{R}_j}\,\widehat{Q}^{(j)}, \qquad j=1,2,3,
  \label{eq:lin_flux}
\end{equation}
where $\widehat{Q}^{(j)}$ denotes the linearised flux.

\subsubsection{Linearised fluxes}

Linearising \eqref{eq:Q_flux_form} about the base state and using
$p_{k,z}^{(0)}|_{\mathrm{base}}=0$ gives
\begin{equation}
  \widehat{Q}^{(j)} =
  \sum_{k=1}^3 A_{jk}(\bar{\mathbf R})\,\widehat{p}_{k,z}^{(0)}
  +
  \sum_{\ell=1}^3
  \frac{\partial G_j}{\partial R_\ell}(\bar{\mathbf R})\,\hat R_\ell.
\end{equation}
Differentiating the linearised forms of
\eqref{eq:p1}--\eqref{eq:p3} yields
\begin{align}
  \widehat{p}_{1,z}^{(0)} &= \frac{1}{\sigma_1 C}
                            \left(\frac{ik\,\hat{R}_1}{\bar{R}_1^2} - i k^3 \hat{R}_1\right), \\
  \widehat{p}_{2,z}^{(0)} &= \widehat{p}_{1,z}^{(0)}
                            - \frac{1}{\sigma_2 C}
                            \left(\frac{ik\,\hat{R}_2}{\bar{R}_2^2} - i k^3 \hat{R}_2\right), \\
  \widehat{p}_{3,z}^{(0)} &= \widehat{p}_{2,z}^{(0)}
                            - \frac{1}{C}
                            \left(\frac{ik\,\hat{R}_3}{\bar{R}_3^2} - i k^3 \hat{R}_3\right).
\end{align}

\subsubsection{Dispersion relation}

Substituting these expressions into \eqref{eq:lin_flux} gives rise to a homogeneous linear system
\begin{equation}
  \mathsf{M}^{LW}(k,s)
  \begin{pmatrix}
    \hat{R}_1\\
    \hat{R}_2\\
    \hat{R}_3
  \end{pmatrix}
  = 0,
  \label{eq:lw_matrix}
\end{equation}
where $\mathsf{M}^{LW}(k,s)$ is a $3\times 3$ matrix whose entries depend algebraically
on $k$, the base radii, and the physical parameters. Non-trivial solutions exist if and only if
\begin{equation}
  \det \mathsf{M}^{LW}(k,s) = 0.
  \label{eq:lw_dispersion}
\end{equation}
This condition yields a cubic polynomial in $s$,
\begin{equation}
  s^3 + a_2(k)\,s^2 + a_1(k)\,s + a_0(k) = 0,
\end{equation}
where the coefficients $a_j(k)$ are explicit algebraic functions of $k$, but are too lengthy to be included here.

\subsection{Comparison with the arbitrary-wavenumber theory}

\begin{figure}
  \centering
  \includegraphics[width=0.75\textwidth]{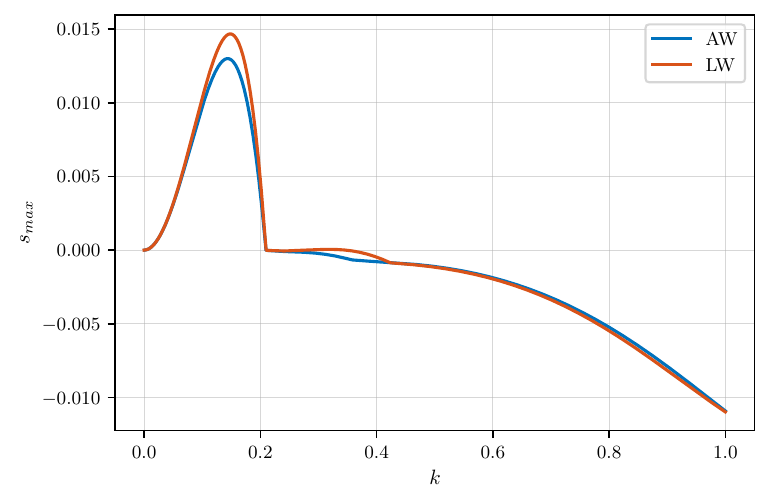}
  \caption{
    Comparison of the growth rate $s$ as a function of wavenumber $k$
    for the dominant mode. The arbitrary-wavenumber results (AW) are shown
    as a solid blue line, while the long-wave approximation (LW) is shown
    as a solid orange line. The parameter values are $a=16.67$,$h_3=1$, $h_2=1$, $h_1=4.67$,  $\sigma_1=0.1$, $\sigma_2=1$, $C=0.4$, $m_1=m_2=1$.
  }
  \label{fig:lw_aw_comparison_case1}
\end{figure}

We compare the dispersion relation obtained from the long-wave model with the arbitrary-wavenumber formulation of \S3 to assess the accuracy and range of validity of the long-wave approximation. Figure~\ref{fig:lw_aw_comparison_case1} shows the growth rate $s_{\max}$ as a function of the wavenumber $k$ for the dominant instability mode.
As expected, the long-wave model agrees closely with the arbitrary-wavenumber
theory in the limit of small $k$. In particular, the two curves are nearly
indistinguishable near $k=0$.

A notable feature of the comparison is the excellent agreement in the cut-off
wavenumber. Both formulations predict a transition from instability to stability
at approximately $k \approx 0.2$, indicating that the long-wave model may provide an accurate prediction of the neutral stability boundary. This agreement is significant given that the long-wave model neglects higher-order axial variations.

At intermediate wavenumbers the long-wave model slightly over predicts the
growth rate, with the largest discrepancy occurring near the peak of the
dispersion curve. Nevertheless, the agreement remains good throughout the
unstable wavenumber range. Beyond the neutral point, the two models remain
qualitatively consistent, although quantitative differences increase as the
assumptions underlying the long-wave approximation become less accurate.

Additional comparisons for other parameter values (not shown) exhibit similar trends. In particular, the long-wave model consistently captures the small-wavenumber behaviour and accurately predicts the cut-off wavenumber. 

\section{Nonlinear evolution}\label{sec:nonlinear_evolution}

We now investigate the nonlinear development of disturbances governed by the
long-wave evolution equations derived in \S4. While the linear stability analysis
predicts the initial growth or decay of perturbations, it does not determine the
ultimate fate of the flow. The central question is whether
unstable disturbances saturate to finite-amplitude states or instead lead to the
formation of singular structures in which one or more layers thin to zero.

To address this, we integrate the evolution equations~\eqref{eq:Rj_flux_final} subject to
periodic boundary conditions on the domain $z\in[-L/2,L/2]$ of length $L=30\pi$.
The initial condition is a small-amplitude cosine perturbation applied to all
three interfaces,
\begin{equation}
  R_i(z,0)=\bar R_i + A_i\cos(k_0 z), \qquad i=1,2,3,
  \label{eq:ic}
\end{equation}
with equal amplitudes $A_1=A_2=A_3=0.01$ and $k_0=4\pi/L$, so that the initial
wavenumber lies within the unstable long-wave band identified in \S3. Spatial
derivatives are approximated by fourth-order central finite differences on a
uniform grid of $N$ points, and the semi-discrete system is advanced by the
method of lines using the SUNDIALS CVODE solver, with adaptive
backward-differentiation time stepping with relative and absolute tolerances of
$10^{-10}$ and $10^{-12}$ respectively. Spatial convergence was
confirmed by successively increasing $N$ until the interface profiles and their
extrema were insensitive to further refinement; the results below use $N=1024$.

Three qualitatively distinct outcomes are observed. In some parameter regimes
the instability saturates, with the system evolving towards a finite-amplitude
travelling wave in which all three layer thicknesses remain bounded away from
zero. In others the evolution leads to rapid, localised thinning and finite-time
breakdown, occurring in one of two ways: closure of the air core, corresponding
to plug formation, or collapse of the intermediate liquid layer. 
To distinguish between these outcomes in the numerical simulations, we adopt
the following classification. The flow is taken to saturate when the interface
extrema approach constant values with all thicknesses bounded away from zero.
Finite-time breakdown is declared when the air-core radius or a layer thickness
first falls below $10^{-4}$: the event is plug formation when it is the minimum
core radius $\min_z R_1$ that vanishes, and middle-layer rupture when it is the
minimum middle-layer thickness $\min_z h_2=\min_z\!\big(R_3-R_2\big)$, with
$\min_z h_1$ and $\min_z h_3$ remaining bounded away from zero. As this threshold is 
approached, the system stiffens and the adaptive CVODE
time step method breaks down, consistent with the onset of a finite-time singularity.
Where the integration could not be continued to the threshold, the outcome was
instead classified from the time histories of $\min_z R_1$, $\min_z h_1$,
$\min_z h_2$ and $\min_z h_3$, by identifying which quantity decreased
rapidly toward zero while the others remained finite.
Simulations reaching neither condition by $t=10^4$ are classified as
saturated. These behaviours are illustrated for representative parameter values
in the following subsections.

\subsection{No plug formation}
\begin{figure}
  \centering
  \includegraphics[width=0.95\textwidth]{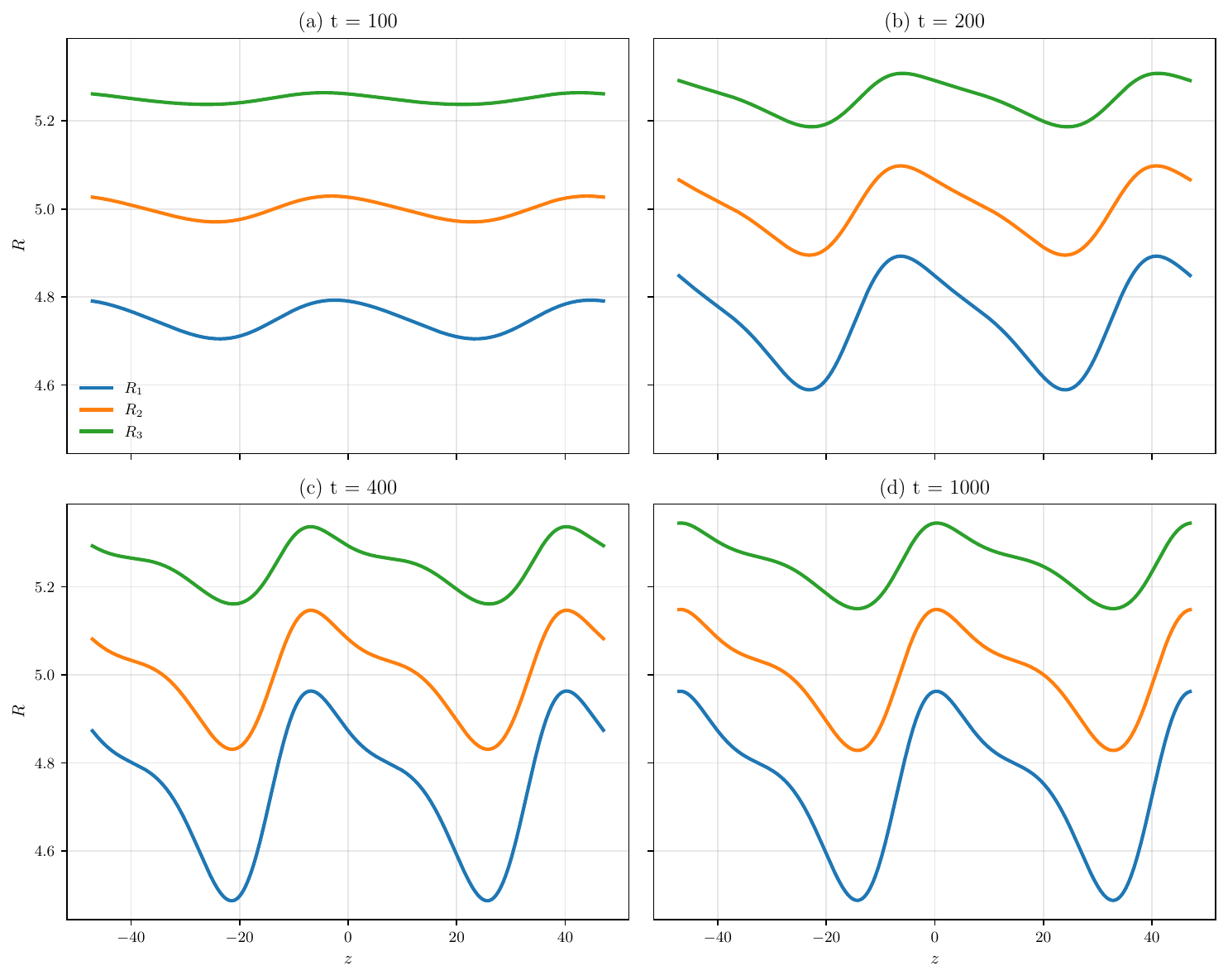}
  \caption{Evolution of the interface profiles in the saturated
travelling-wave regime. The panels correspond to
(a) $t=100$, (b) $t=200$, (c) $t=400$, and (d) $t=1000$.
The parameter values are
$m_1=0.5$, $m_2=0.5$, $a=6.25$,
$\bar h_1=\bar h_2=0.25$,
$\bar h_3=1$, $\sigma_1=0.1$,
$\sigma_2=1$, $L=30\pi$, and $C=0.4$.}
  \label{fig:no_plug_profiles}
\end{figure}

\begin{figure}
  \centering
  \includegraphics[width=0.95\textwidth]{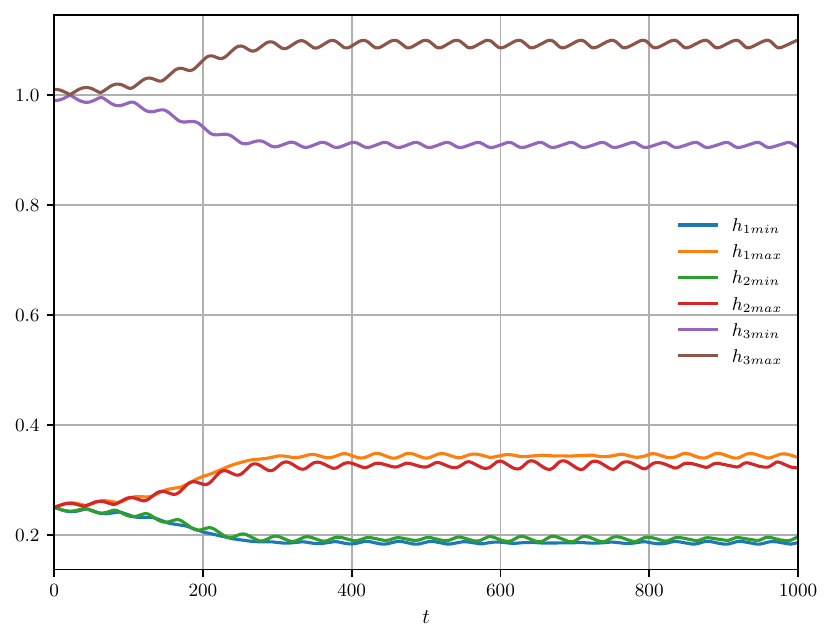}
  \caption{Time evolution of the minimum and maximum layer thicknesses for the same parameters 
  as figure~\ref{fig:no_plug_profiles}.}
  \label{fig:no_plug_minmax}
\end{figure}

We first consider a parameter regime in which the instability does
not lead to rupture but instead saturates at finite amplitude.
The parameter values are $m_1=0.5$, $m_2=0.5$, $a=6.25$, $\bar h_1=\bar h_2=0.25$, $\bar h_3=1$, $\sigma_1=0.1$, $\sigma_2=1$, $L=30\pi$, and $C=0.4$.
The nonlinear evolution for this parameter set is shown in
figures~\ref{fig:no_plug_profiles} and \ref{fig:no_plug_minmax}. Following an initial phase of exponential growth, the interface deformations increase smoothly before saturating at finite amplitude. By approximately $t=400$, the solution has approached has essentially converged to a travelling-wave solution whose shape changes very little thereafter. These travelling-wave solutions are investigated further in \S5.3.

\subsection{Finite-time rupture}

We next consider parameter regimes in which the nonlinear evolution does not
saturate, but instead leads to the rapid thinning of one of the layers. In the
three-layer system, finite-time breakdown can occur in more than one way. We
first describe a plug-forming case, in which the air core closes, and then
present a second example in which the middle liquid layer ruptures.

\subsubsection{Core collapse and plug formation}

\begin{figure}
  \centering
  \includegraphics[width=0.95\textwidth]{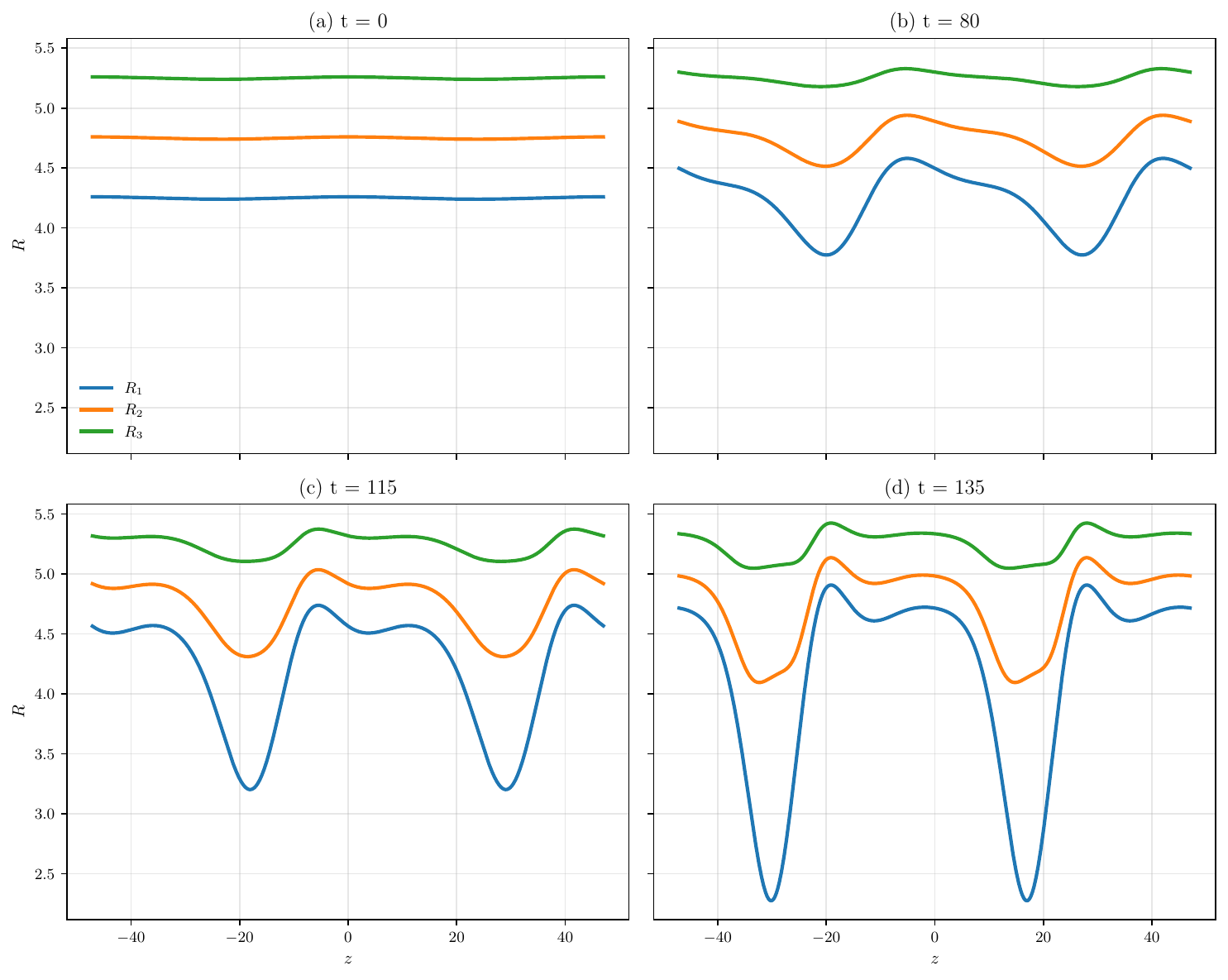}
  \caption{Evolution of the interface profiles in the plug-forming regime at selected times. The parameter values are
$m_1=0.5$, $m_2=0.5$, $a=6.25$, $\bar h_1=\bar h_2=0.5$, $\bar h_3=1$, $\sigma_1=0.1$, $\sigma_2=1$, $L=30\pi$, and $C=0.4$.}
  \label{fig:plug_profiles}
\end{figure}

Figure~\ref{fig:plug_profiles} shows the evolution of the interface profiles
for a parameter regime in which the nonlinear dynamics lead to plug formation. The panels 
correspond to times (a) $t=0$, (b) $t=80$, (c) $t=115$, and (d) $t=135$. The parameters are the same as in the no-rupture case, except that $\bar h_1=\bar h_2=0.5$. Starting from a nearly uniform state, the disturbances initially grow in a manner consistent with the linear instability. At intermediate times, as shown in figure~\ref{fig:plug_profiles}(b), the deformation of each interface remains smooth and even across the domain. 
As the evolution proceeds, the deformation becomes increasingly
localised. By $t=115$, figure~\ref{fig:plug_profiles}(c), the inner
interface develops pronounced troughs, while the middle and outer
interfaces undergo comparatively smaller deformations. This localisation
intensifies further at later times, as shown in
figure~\ref{fig:plug_profiles}(d).

\begin{figure}
  \centering
  \includegraphics[width=0.75\textwidth]{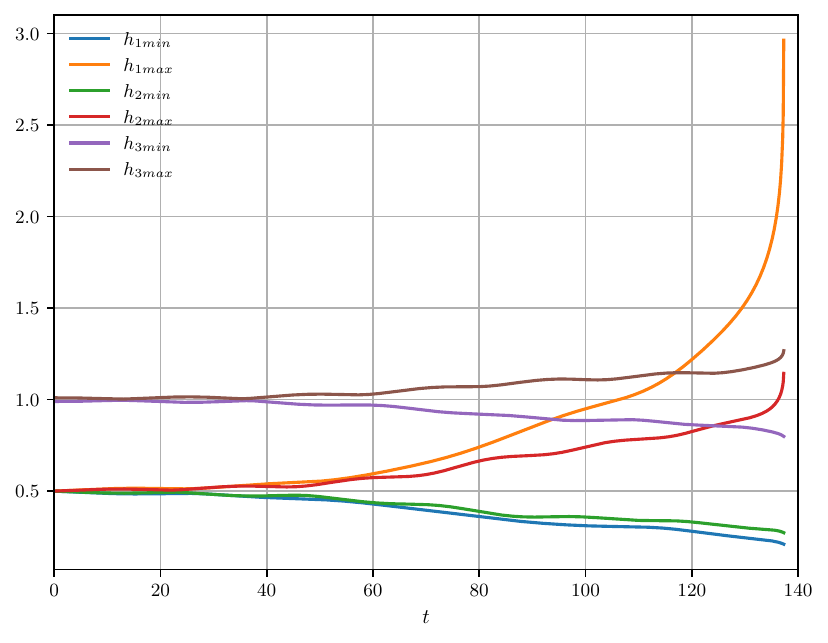}
  \caption{Time evolution of the minimum and maximum layer thicknesses in the plug-forming regime. The parameter values are identical to those in the previous figure.}
  \label{fig:plug_minmax}
\end{figure}

The temporal evolution of the extrema, displayed in
figure~\ref{fig:plug_minmax}, provides a complementary view of this process.
The minimum inner-layer thickness decreases monotonically, while the maxima
increase rapidly in time. In particular, the growth accelerates at later times,
indicating the absence of nonlinear saturation. Together with the rapid inward
motion of the inner interface observed in figure~\ref{fig:plug_profiles}, this
behaviour is consistent with closure of the air core.

This combined behaviour, progressive thinning, rapid amplification of the
interface extrema, and collapse of the air-core radius, is characteristic of
the onset of a finite-time singularity. We identify this regime with
plug formation.

\subsubsection{Middle-layer rupture}

\begin{figure}
  \centering
  \includegraphics[width=0.95\textwidth]{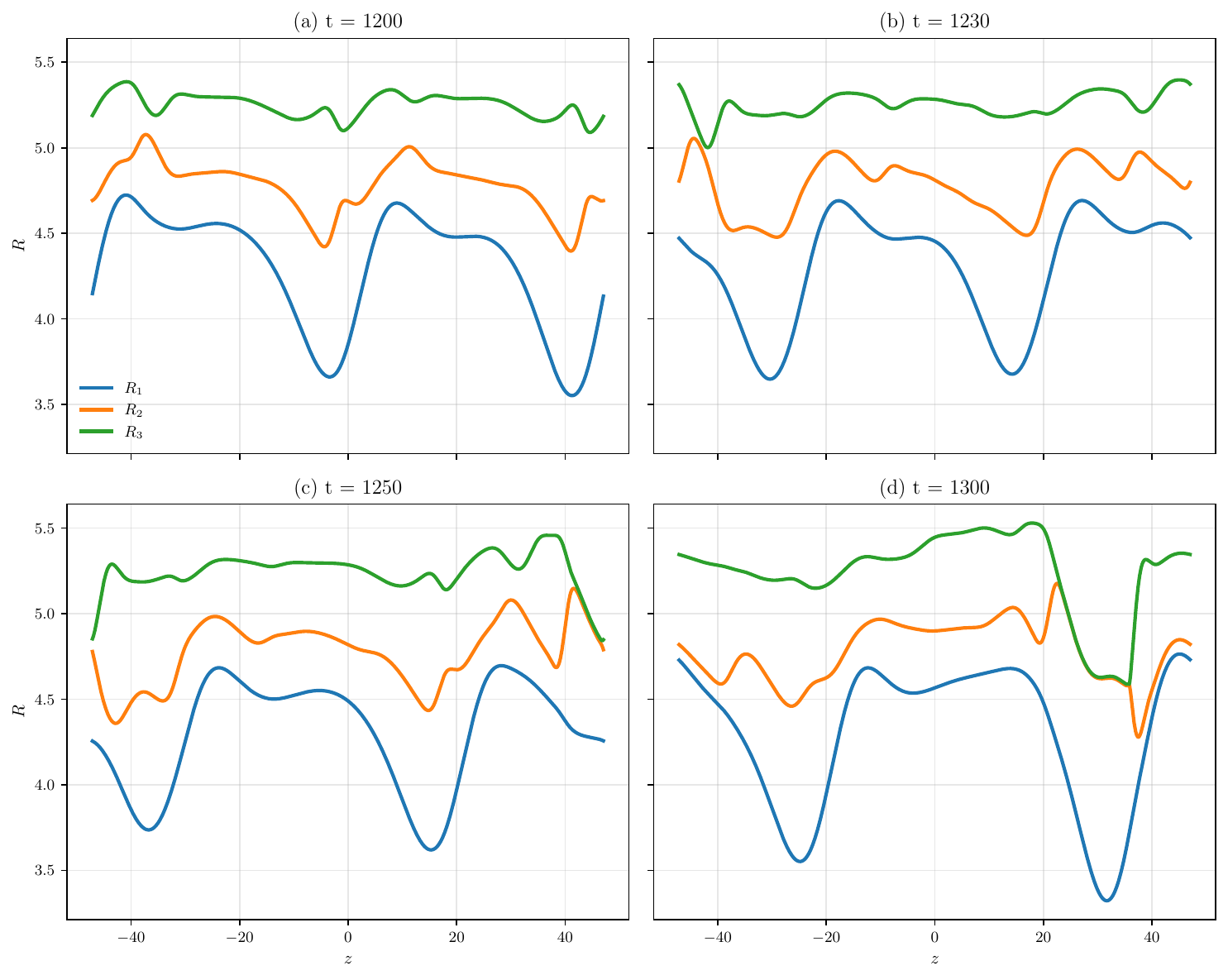}
  \caption{Evolution of the interface profiles in the middle-layer rupture
    regime. The panels correspond to times (a) $t=1200$, (b) $t=1230$,
    (c) $t=1250$, and (d) $t=1300$. In contrast to the plug-forming case,
    the air-core radius remains bounded away from zero while the middle
    layer becomes strongly localised and thins near rupture.	The parameter values are 
    $m_1=0.5$, $m_2=2$, $a=6.25$, $\bar h_1=\bar h_2=0.469$,
    $\bar h_3=1$, $\sigma_1=0.1$, $\sigma_2=1$, $L=30\pi$ and $C=0.4$.}
  \label{fig:midlayer_rupture_profiles}
\end{figure}

\begin{figure}
  \centering
  \includegraphics[width=0.75\textwidth]{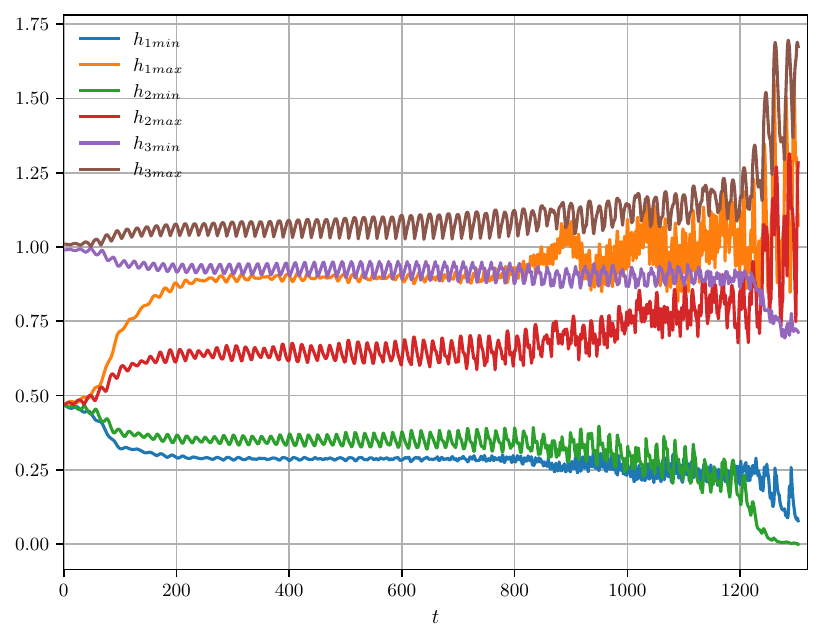}
  \caption{Time evolution of the minimum layer thicknesses in the
    middle-layer rupture regime. The middle-layer thickness
    $\min h_2$ decreases rapidly to zero, while $\min h_1$ and
    $\min h_3$ remain bounded away from zero, confirming that rupture
    occurs through collapse of the intermediate layer rather than
    closure of the air core.}
  \label{fig:midlayer_rupture_min}
\end{figure}

A representative example of the second type of finite-time breakdown is shown in figure~\ref{fig:midlayer_rupture_profiles}. In this case the air core remains
open, but the middle liquid layer thins locally as the interfaces $R_2$ and
$R_3$ approach each other. The early profiles are smooth and resemble a finite-amplitude travelling wave. At later times, however, the deformation becomes increasingly localised, and the coupled evolution of the three interfaces drives the progressive thinning of the middle layer. The inner interface remains separated from the tube axis, while the gap between $R_2$ and $R_3$ becomes increasingly small in localised regions. This behaviour is quantified in figure~\ref{fig:midlayer_rupture_min}, which shows the evolution of the minimum layer thicknesses. The middle-layer thickness $\min (h_2)$ decreases rapidly towards zero, whereas $\min (h_1)$ and $\min (h_3)$ remain strictly positive throughout the evolution.

\begin{figure}
  \centering
  \includegraphics[width=0.75\textwidth]{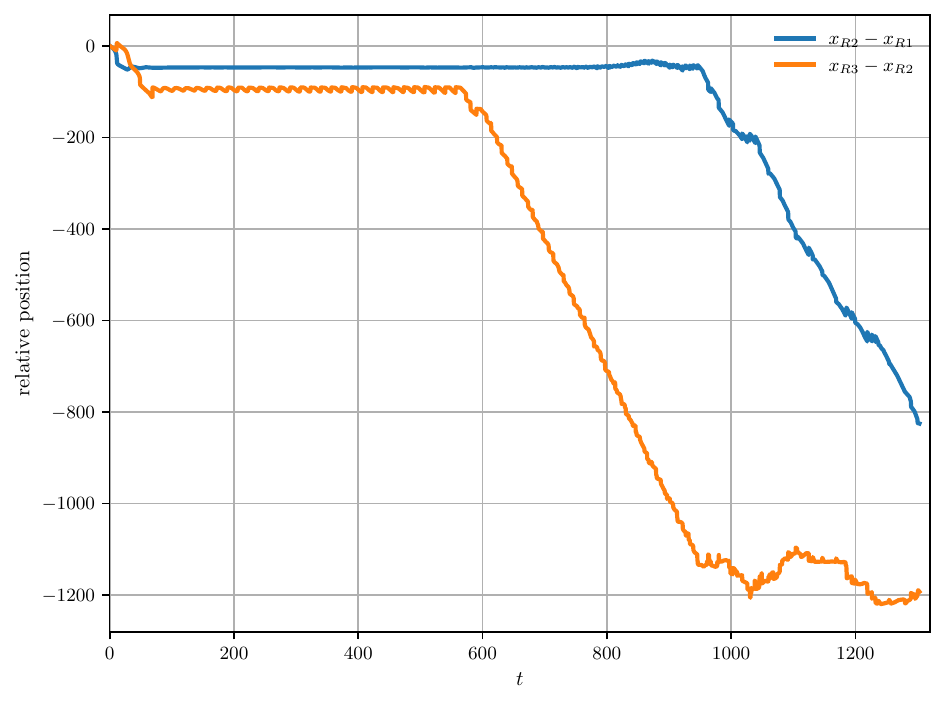}
  \caption{Relative displacement of the interface trough locations as a function of time, showing $x_{R_2}-x_{R_1}$ and $x_{R_3}-x_{R_2}$.
    After an initial transient, both quantities exhibit sustained drift,
    indicating that the interfaces propagate at different effective speeds.
    The separation between $R_2$ and $R_3$ remains comparatively smaller than
    that between $R_1$ and $R_2$, consistent with the strong coupling of the
    outer two interfaces and the progressive thinning of the middle layer.}
  \label{fig:midlayer_drift}
\end{figure}

Tracking the motion of the interface troughs provides additional insight,
as shown in figure~\ref{fig:midlayer_drift}. After an initial transient,
both relative displacements exhibit sustained drift, indicating that the
three interfaces propagate with different effective speeds. Throughout the
evolution, the separation between the troughs of $R_2$ and $R_3$ remains
considerably smaller than that between $R_1$ and $R_2$, demonstrating the
strong coupling of the outer two interfaces. This differential motion
progressively reduces the local thickness of the middle layer and ultimately
leads to rupture.

To better understand the saturated states observed in the nonlinear
simulations, we now compute travelling-wave solutions of the long-wave model.

\subsection{Travelling-wave formulation and numerical continuation}

We seek nonlinear travelling-wave solutions of the long-wave evolution equations
in the form
\begin{equation}
  R_j(z,t)=\mathcal R_j(Z), \qquad Z=z-ct, \qquad j=1,2,3,
\end{equation}
where \(c\) is the wave speed and the profiles \(\mathcal R_j\) are periodic with period \(L\). Substituting this ansatz into the long-wave evolution equations \eqref{eq:Rj_flux_final} gives
\begin{equation}
  -c\,\mathcal R_j'
  =
  \frac{1}{\mathcal R_j}\frac{dQ^{(j)}}{dZ},
  \qquad j=1,2,3,
\end{equation}
which, upon multiplication by \(\mathcal R_j\) and integration in \(Z\), yields
\begin{equation}
  Q^{(j)}+\frac{c}{2}\mathcal R_j^{\,2}=K_j,
  \qquad j=1,2,3,
  \label{eq:TW_integrated_flux}
\end{equation}
where \(K_j\) are constants of integration. The travelling-wave profiles therefore satisfy the coupled 
system
\begin{equation}
  \sum_{k=1}^3
  A_{jk}(\mathcal R)\,P_k
  +
  G_j(\mathcal R)
  +
  \frac{c}{2}\mathcal R_j^{\,2}
  -K_j
  =0,
  \qquad j=1,2,3,
  \label{eq:TW_compact}
\end{equation}
where \(P_k = p^{(0)}_{k,z}\big|_{R_j=\mathcal R_j}\). Using the expressions for the pressure 
gradients, (\ref{eq:TW_compact}) yields a system of three coupled
third-order ordinary differential equations for \(\mathcal R_1,\mathcal R_2,\mathcal R_3\).
Periodic travelling waves satisfy
\begin{equation}
  \mathcal R_j(0)=\mathcal R_j(L), \qquad j=1,2,3,
\end{equation}
together with constraints fixing the mean cross-sectional areas of each layer.
A phase condition is imposed to remove translational invariance. Explicit forms of the governing equations, together with details of the numerical approach, are provided in Appendix~B.

The formulation above defines a nonlinear periodic boundary-value problem
for travelling-wave solutions parameterised by the mean layer thicknesses, viscosity ratios and surface tension ratios. In the computations presented here we restrict attention to the representative family $\bar h_2=\bar h_1$,
so that the two inner liquid layers have the same mean thickness and
the branch may be parameterised by a single thickness variable
\(\bar h_1\).
This choice reduces the dimensionality of the parameter space while allowing
the effects of mean film thickness and viscosity ratio $m_2=\mu_3/\mu_2$ to
be investigated systematically, although other thickness ratios may modify
the quantitative details of the branch structure.
\begin{figure}
  \centering
  \includegraphics[width=\textwidth]{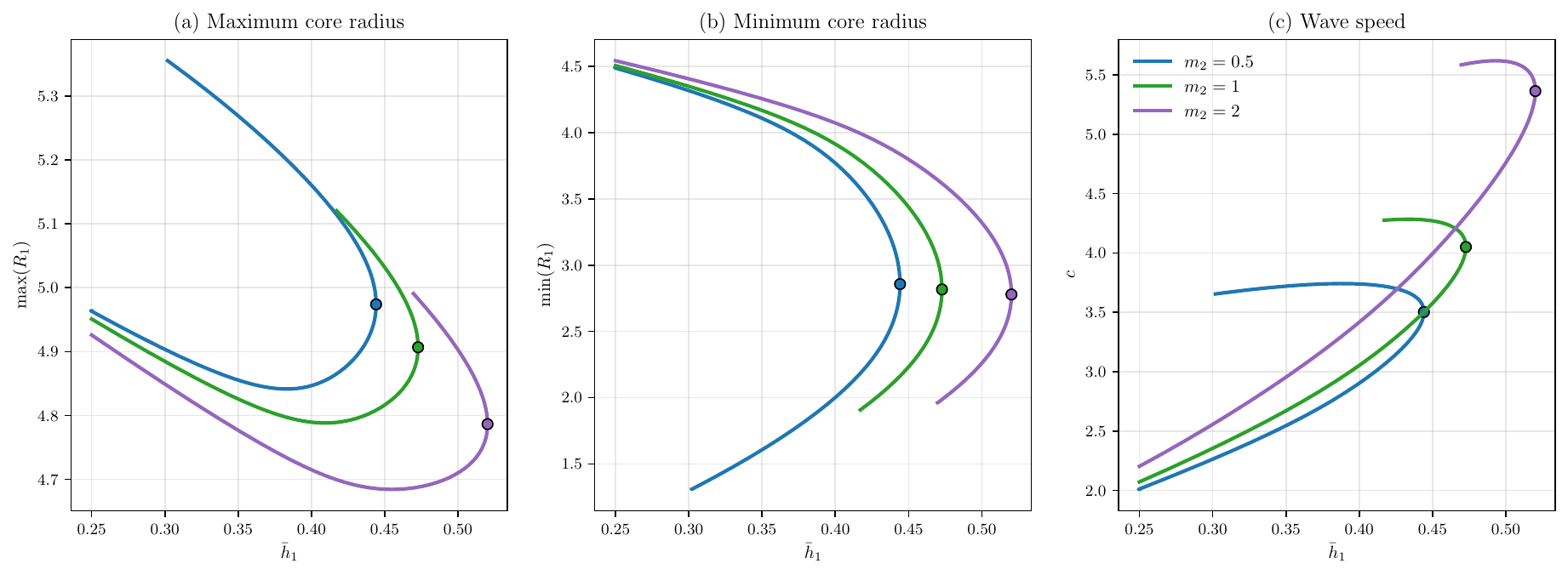}
  \caption{Travelling-wave branches obtained by continuation in $\bar h_1$
    with $\bar h_2=\bar h_1$, for different values of $m_2$.
    Panels show (a) $\max(R_1)$, (b) $\min(R_1)$, and (c) wave speed $c$
    as functions of $\bar h_1$. Filled circles denote the limit points
    on the corresponding branches.}
  \label{fig:tw_branch_m2}
\end{figure}

Figure~\ref{fig:tw_branch_m2} shows the travelling-wave solution branches
obtained by numerical continuation for
$a=6.25$, $\bar h_3=1$, $C=0.4$, $\sigma_1=0.1$,
$\sigma_2=1$, and $L=30\pi$, using $\bar h_1$ as the continuation
parameter for several values of $m_2$. Panels (a)--(c) display the
corresponding variations of $\max(R_1)$, $\min(R_1)$, and the wave
speed $c$, respectively. For each value of $m_2$, the branch exists
over a finite interval of $\bar h_1$ and folds at a limit point (filled
circles). Consequently, for values of $\bar h_1$ below the fold, two
distinct travelling-wave solutions coexist on the same branch
projection. The quantities $\max(R_1)$ and $\min(R_1)$ provide
additional measures of the wave shape. The former characterises the
largest outward displacement of the core interface, whereas the latter
measures the strongest constriction of the air core. As $m_2$
increases, the branches extend to larger values of $\bar h_1$, while
the corresponding variation of the wave speed is shown in panel (c).

\begin{figure}
  \centering
  \includegraphics[width=\textwidth]{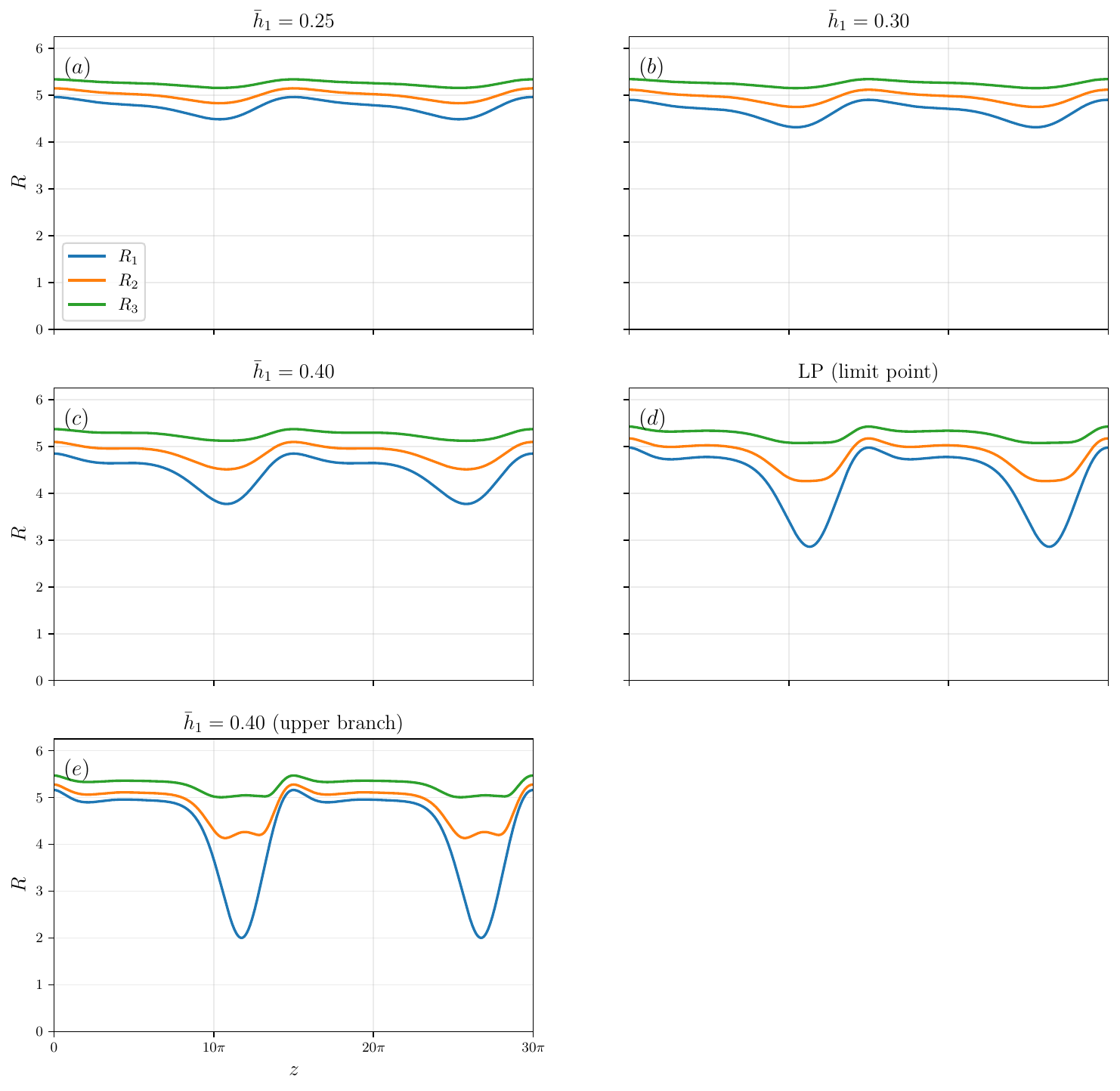}
  \caption{Representative travelling-wave profiles along the branch.
    Panels correspond to increasing $\bar h_1$, including a configuration
    near the limit point (LP).}
  \label{fig:tw_profiles}
\end{figure}

Representative travelling-wave profiles along the branch are displayed
in figure~\ref{fig:tw_profiles}. For smaller $\bar h_1$, the waves are
of small amplitude and nearly sinusoidal. As $\bar h_1$ increases, the profiles steepen and develop pronounced localised troughs in $R_1$, indicating strong
interfacial deformation. Near the limit point, the travelling waves develop strongly localised troughs and lose their nearly sinusoidal character, perhaps reflecting the importance of nonlinear effects. Panel (e) shows the second travelling-wave solution at $\bar h_1=0.40$ on the upper branch. Compared with the solution on the lower branch shown in panel (c), the wave exhibits substantially deeper constriction of the air core and stronger localisation.

\begin{figure}
  \centering
  \includegraphics[width=\textwidth]{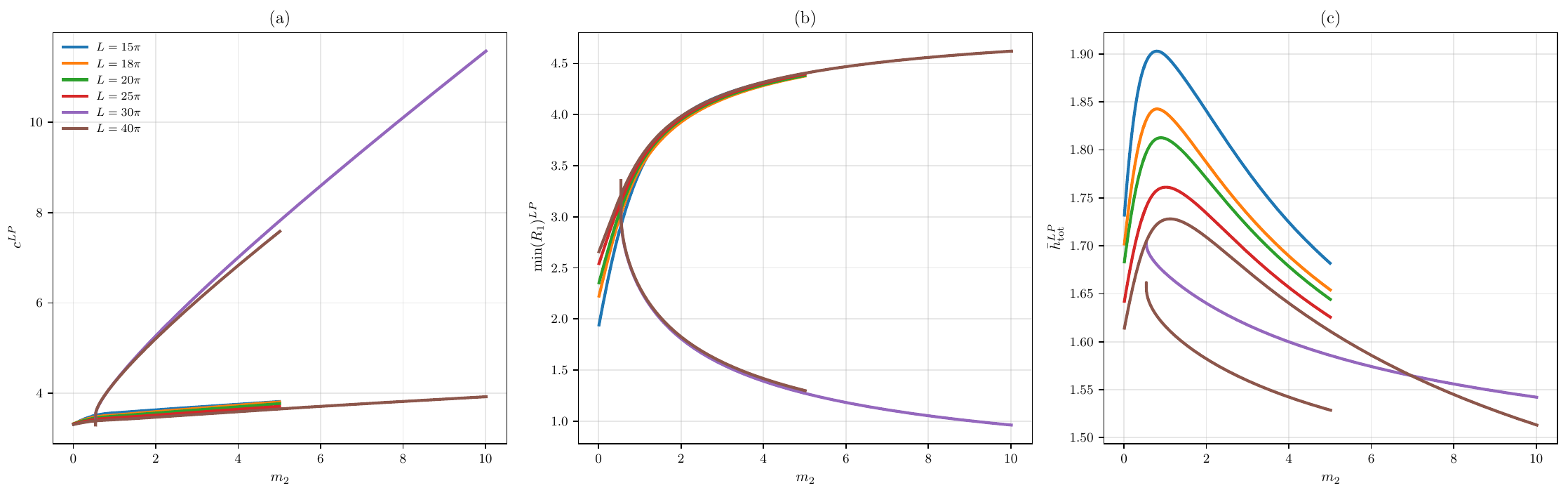}
  \caption{
    Dependence of limit-point quantities on the viscosity ratio $m_2$
    for travelling waves of different periods $L$.
    Each curve is obtained by locating the limit point of a travelling-wave
    branch continued in $\bar h_1$ (with $\bar h_2=\bar h_1$) at fixed
    $m_2$. Panels show (a) the limit-point wave speed $c^{LP}$,
    (b) the limit-point minimum core radius $\min(R_1)^{LP}$, and
    (c) the corresponding total thickness
    $\bar h_{\mathrm{tot}}^{LP}=1+2\bar h_1^{LP}$.
    For $L=30\pi$ a second limit-point branch is also found, producing the
    lower curve visible in panels (b) and (c).
  }
  \label{fig:tw_LP_m2}
\end{figure}

Figure~\ref{fig:tw_LP_m2} demonstrates the dependence of the
travelling-wave limit points on both the viscosity ratio $m_2$
and the wave period $L$. For each pair $(L,m_2)$, a travelling-wave branch was continued in $\bar h_1$ (with $\bar h_2=\bar h_1$), and the corresponding limit point was recorded. Thus figure~\ref{fig:tw_LP_m2} does not show travelling-wave branches themselves, but rather the locus of their turning points in parameter space.

The motivation for tracking these limit points is that turning points of
travelling-wave branches have been used as indicators of critical film
thicknesses for plug formation in related long-wave models.
In particular, \citet{Ogrosky2021} used the turning point of a
free-surface travelling-wave branch as a proxy for the critical
thickness beyond which plugs are expected to form.
We interpret $\bar h_{\mathrm{tot}}^{LP}$ as an analogous
indicator of the critical film thickness for the present three-layer
system, while recognising that this must ultimately be
validated by the time-dependent simulations presented below.

Several trends are apparent.
For moderate periods, $15\pi \le L \le 25\pi$, the limit-point loci vary
smoothly with $L$ and remain relatively close to one another.
At fixed $m_2$, increasing the period generally decreases
$\bar h_{\mathrm{tot}}^{LP}$ and increases $\min(R_1)^{LP}$,
indicating that longer waves can persist to smaller film thicknesses
before reaching a turning point.

A qualitatively different behaviour is observed for $L=30\pi$, where a
second limit-point branch is detected.
This branch is characterised by substantially smaller values of
$\bar h_{\mathrm{tot}}^{LP}$ and $\min(R_1)^{LP}$ and extends to much
larger values of $m_2$ than the primary branch. For $L=40\pi$, only a single limit-point branch was identified over the range of parameters considered.
Its continuation extends to larger values of $m_2$ than for the shorter
periods, while yielding the smallest values of $\bar h_{\mathrm{tot}}^{LP}$ among the branches shown. The emergence of an additional limit-point branch near $L=30\pi$ suggests that the travelling-wave solution structure undergoes a
qualitative reorganisation as the period increases.

\begin{figure}
  \centering
  \includegraphics[width=\textwidth]{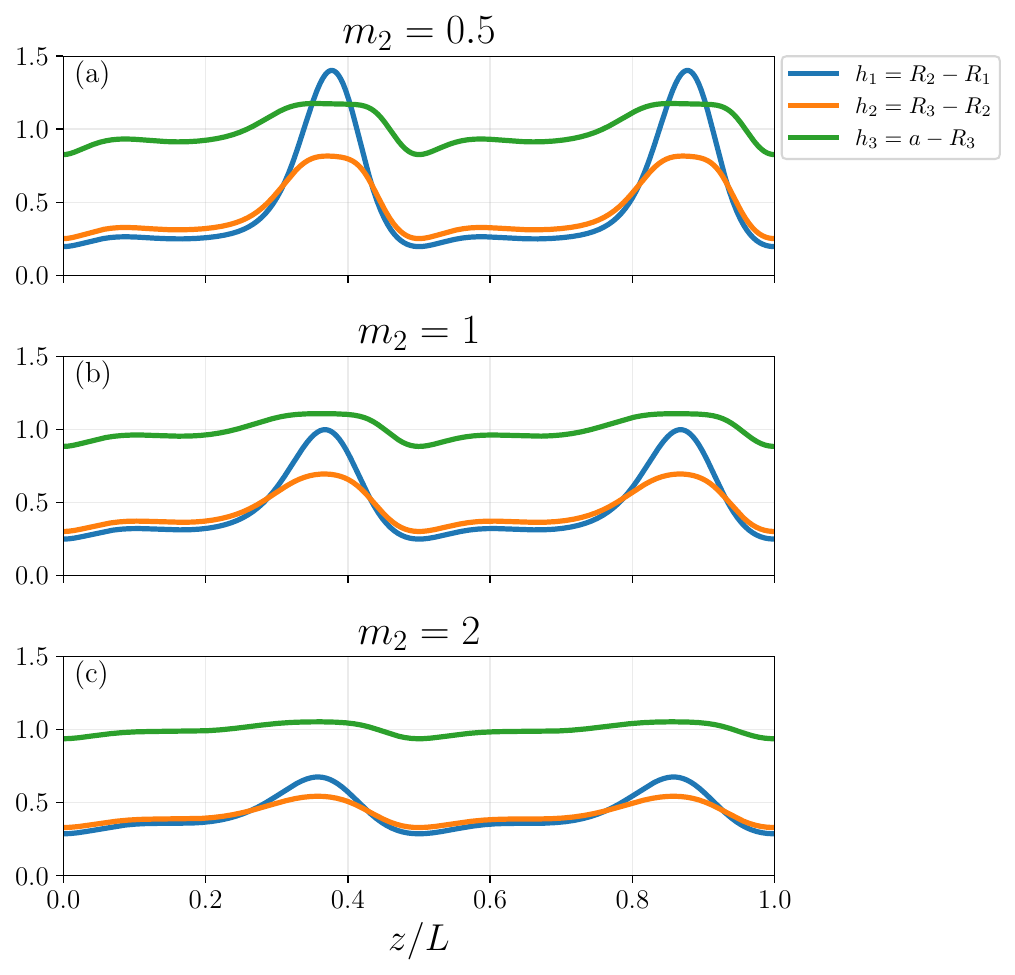}
  \caption{Layer-thickness profiles of travelling-wave solutions near the
    limit point for $m_2=0.5$, $1$, and $2$ with $m_1=0.5$. The thicknesses are defined by
    $h_1=R_2-R_1$, $h_2=R_3-R_2$, and $h_3=a-R_3$. As $m_2$ increases, the deformation of the inner layer becomes
    progressively weaker, while the outer layer remains comparatively
    uniform. The strongest localisation occurs in the inner layer for
    smaller $m_2$, consistent with the plug-forming dynamics observed
    in the time-dependent simulations.}
  \label{fig:LP_thickness_profiles}
\end{figure}

Figure~\ref{fig:LP_thickness_profiles} illustrates how the layer-thickness profiles of travelling-wave solutions near the limit point evolve as the viscosity ratio $m_2$ increases. For $m_2=0.5$, the deformation is strongly concentrated in the inner layer, with pronounced peaks and troughs in $h_1$ corresponding to severe constriction of the air core. Increasing $m_2$ progressively reduces the amplitude of these variations, while the middle layer undergoes more modest changes and the outer layer remains nearly uniform throughout. This systematic redistribution of the deformation is consistent with the time-dependent simulations, in which smaller $m_2$ promotes plug formation through collapse of the air core, whereas larger $m_2$ favours rupture of the middle layer.

\begin{figure}
  \centering
  \includegraphics[width=0.8\textwidth]{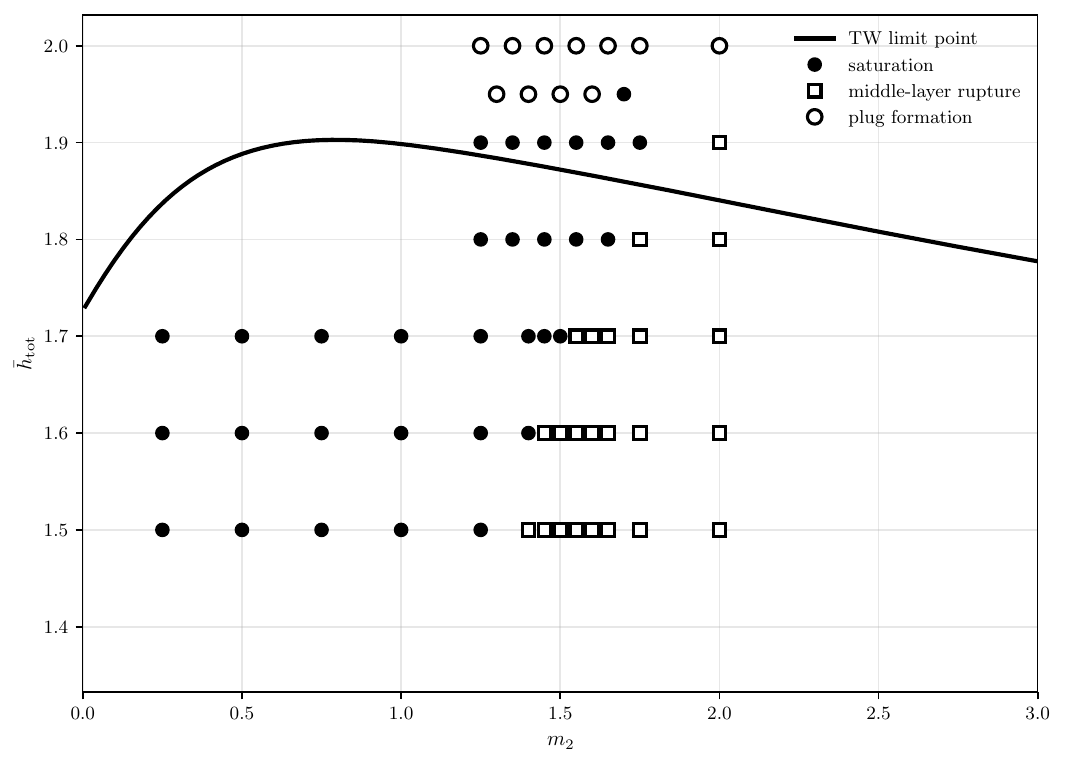}
  \caption{Comparison between the travelling-wave limit-point locus and
    the outcomes of time-dependent simulations for $m_1=0.5$.
    Filled circles denote cases that evolve towards finite-amplitude
    saturated states, open circles correspond to plug formation through
    closure of the air core, and open squares indicate rupture of the
    middle layer.
    The solid curve shows the total thickness at the travelling-wave
    limit point,
    $\bar h_{\mathrm{tot}}^{LP}=1+2\bar h_1^{LP}$,
    as a function of $m_2$.}
  \label{fig:LP_outcomes}
\end{figure}

Figure~\ref{fig:LP_outcomes} compares the outcomes of the time-dependent
simulations with the travelling-wave limit-point locus as the viscosity
ratio $m_2$ is varied. As in the corresponding two-layer
problem, the limit point of the free-surface (air-core) branch seems to serve as a proxy
for the critical thickness for plug formation: for smaller $m_2$, where
breakdown occurs through closure of the air core, cases near and above the locus
evolve towards plug formation, consistent with the strong inner-layer
deformation seen near the limit point (figure~\ref{fig:LP_thickness_profiles}).
The locus does not, however, mark a sharp boundary between saturation and
rupture. For larger $m_2$, breakdown instead occurs predominantly through
collapse of the middle layer, a route that involves the two internal interfaces
and is not represented by the free-surface branch, and these cases
(open squares) depart from the locus; a few saturated states are also found
below it. The limit-point locus remains a useful indicator of the
air-core-closure threshold, as in the corresponding two-layer problem.
However, the full rupture behaviour of the three-layer system,
including the transition to middle-layer rupture as $m_2$ increases,
cannot be inferred from the free-surface limit point alone.

\section{Conclusion} \label{sec:conclusion}

We have developed and analysed a model for the axisymmetric evolution of
three immiscible viscous films lining the interior of a vertical cylindrical
tube containing a passive air core. Starting from the axisymmetric
Navier--Stokes equations, we first performed an arbitrary-wavenumber
linear stability analysis of the full Stokes system. We then derived a reduced long-wave model governing the nonlinear evolution of the three interfaces in the limit where the characteristic
layer thickness is small compared with the axial wavelength. The resulting
long-wave equations were used to investigate both the linear and nonlinear
dynamics of the system, including travelling-wave solutions and rupture
behaviour.

The arbitrary-wavenumber analysis reveals a persistent long-wave
instability together with a finite-wavenumber structure that depends
sensitively on the layer thicknesses, viscosity ratios, and surface
tension ratios. Variations in these parameters alter the degree of
interfacial coupling, producing changes in the unstable wavenumber
spectrum, mode switching, and the suppression of short-wave
disturbances. The long-wave model accurately reproduces the
small-wavenumber behaviour, including the cut-off wavenumber, and
provides an effective reduced description of the instability.

Results of the numerical integration of the nonlinear evolution equations indicate three qualitatively distinct outcomes. While saturated travelling waves and air-core plug formation are observed (consistent with single- and two-layer studies), we identify a third regime unique to the three-layer configuration: the rupture of the intermediate liquid layer. 
This occurs while the air core remains open, demonstrating that a second internal interface creates competing routes to finite-time breakdown. Numerical continuation of travelling-wave 
branches, together with
time-dependent simulations, shows that the dominant rupture mechanism
depends strongly on the viscosity ratio $m_2$.  We find that 
smaller $m_2$ values typically lead to air-core closure, whereas larger $m_2$ values trigger 
intermediate layer rupture. The locus of travelling-wave limit points provides a useful framework
for interpreting the transition between different nonlinear rupture
mechanisms, although it does not constitute a sharp stability boundary.

Several directions for future work suggest themselves. The equal-density assumption adopted here could be relaxed to incorporate density stratification, which would introduce additional buoyancy-driven mechanisms. The near-rupture dynamics associated with the intermediate-layer collapse identified here
remain largely unexplored. In contrast to the more familiar rupture mechanisms in two-layer thin-film and core--annular flows, the present system permits the localised extinction of the middle layer while both neighbouring layers remain finite. The numerical results suggest the development of strongly localised thinning regions, raising the possibility of self-similar
rupture dynamics analogous to those arising in classical thin-film rupture problems. Determining the asymptotic structure of the thinning profile, the associated scaling laws for the minimum middle-layer thickness, and the extent to which existing rupture theories for lubrication-type free-surface flows can be adapted to this three-layer geometry remain open problems. It would also be of interest to determine whether the rupture exhibits universal
similarity exponents or whether the interaction of the two deformable interfaces introduces qualitatively new singular behaviour.
The present results demonstrate that introducing a third deformable
liquid layer does not merely increase the complexity of the flow, but
fundamentally changes its nonlinear dynamics by creating new routes to
rupture that are absent in one- and two-layer cylindrical films. These
findings provide a framework for understanding the interplay of
multiple deformable interfaces in confined geometries relevant to
multiphase transport.

\begin{bmhead}[Acknowledgements]
  This work was carried out in partial
  fulfilment of the requirements for the Ph.D. degree of Awa Traore at
  the University of Alabama.
  \end{bmhead}

\begin{bmhead}[Funding]
  This research received no specific grant from any funding agency,
  commercial or not-for-profit sectors.
  \end{bmhead}

  \begin{bmhead}[Declaration of interests]
    The authors report no conflict of interest.
    \end{bmhead}

    \begin{bmhead}[Data availability statement]
      The data and computer codes
  supporting the findings of this study, including the
  figure-generation scripts and a \texttt{requirements.txt} file, are
  available from the corresponding author upon reasonable request.
    \end{bmhead}

\begin{appen}

\section{Base-state velocity coefficients}
\label{app:base_coeffs}

The base-state velocities are
\[
  \bar w_j(r)=-{m_j}{4}r^2+a_j\ln r+b_j,
  \qquad j=1,2,3.
\]
The coefficients $a_j$ and $b_j$ are determined from the
shear-free condition at the inner interface,
continuity of velocity and tangential stress at the two liquid-liquid
interfaces, and the no-slip condition at the wall:
\begin{align}
  \bar w_{1,r}(\bar R_1)&=0,\\
  \bar w_1(\bar R_2)&=\bar w_2(\bar R_2),\\
  m_1^{-1}\bar w_{1,r}(\bar R_2)&=
                                  m_2^{-1}\bar w_{2,r}(\bar R_2),\\
  \bar w_2(\bar R_3)&=\bar w_3(\bar R_3),\\
  m_2^{-1}\bar w_{2,r}(\bar R_3)&=
                                  \bar w_{3,r}(\bar R_3),\\
  \bar w_3(a)&=0.
\end{align}

\section{Flux coefficients}
\label{app:flux_coeffs}

In this appendix we record the coefficient functions appearing in the flux representation
of the long-wave evolution equations. As shown in \S4.1, the cumulative fluxes may be written as
\begin{equation}
  Q^{(j)}=\sum_{k=1}^3 A_{jk}(R_1,R_2,R_3)\,p_{k,z}^{(0)}+G_j(R_1,R_2,R_3),
  \qquad j=1,2,3,
\end{equation}
where
\begin{equation}
  Q^{(3)}:=Q_3,\qquad   Q^{(2)}:=Q_2+Q_3,\qquad Q^{(1)}:=Q_1+Q_2+Q_3.
\end{equation}
Here the coefficients \(A_{jk}\) multiply the leading-order pressure gradients, while \(G_j\) denote the contributions from gravitational drainage. We write
\begin{equation}
  Q^{(1)}=A_{11}p_{1,z}^{(0)}+A_{12}p_{2,z}^{(0)}+A_{13}p_{3,z}^{(0)}+G_1.
\end{equation}
The coefficient functions are
\begin{align}
  A_{11} &= \frac{1}{16}\Bigg[
           2m_2\left(R_2^4+R_1^2R_3^2-R_1^2R_2^2-R_2^2R_3^2
           +2R_1^2R_2^2\ln\frac{R_3}{R_2}-2R_1^4\ln\frac{R_3}{R_2}\right)
           \notag\\
         &\qquad
           -m_1\left(3R_1^4+R_2^4-4R_1^2R_2^2+4R_1^4\ln\frac{R_2}{R_1}\right)
           \notag\\
         &\qquad
           +2(R_1^2-R_2^2)\left(a^2-R_3^2-2R_1^2\ln\frac{a}{R_3}\right)
           \Bigg], \\
  A_{12} &= \frac{m_2 R_1^2}{8}\left[2R_2^2\ln\frac{R_2}{R_3}
           + R_3^2 - R_2^2\right]
           + \frac{R_1^2(R_3^2-R_2^2)}{4}\ln\frac{a}{R_3}
           - \frac{m_2(R_2^2-R_3^2)^2}{16}
           + \frac{(R_2^2-R_3^2)(a^2-R_3^2)}{8}, \\
  A_{13} &= \frac{1}{16}\left[
           -a^4-R_3^4+2a^2R_1^2+2a^2R_3^2-2R_1^2R_3^2
           -4R_1^2R_3^2\ln\frac{a}{R_3}
           \right], \\
  G_1 &= \frac{R_1^4}{4}\left[\frac{3m_1}{4}
	+ m_1\ln\frac{R_2}{R_1}
	+ m_2\ln\frac{R_3}{R_2}
	+ \ln\frac{a}{R_3}\right]
	\notag\\
         &\qquad
           + \frac{R_1^2}{4}\left[m_2(R_2^2-R_3^2) - m_1 R_2^2 + R_3^2 - a^2\right]
           \notag\\
         &\qquad
           + \frac{1}{16}\left[m_1 R_2^4 + m_2(R_3^4 - R_2^4) + a^4 - R_3^4\right].
\end{align}
We next write
\begin{equation}
  Q^{(2)}=A_{21}p_{1,z}^{(0)}+A_{22}p_{2,z}^{(0)}+A_{23}p_{3,z}^{(0)}+G_2.
\end{equation}
The corresponding coefficients are
\begin{align}
  A_{21} &= \frac{1}{8}\Bigg[
           m_2\left(R_2^4+R_1^2R_3^2-R_1^2R_2^2-R_2^2R_3^2
           +2R_1^2R_2^2\ln\frac{R_2}{R_3}-2R_2^4\ln\frac{R_2}{R_3}\right)
           \notag\\
         &\qquad
           +(R_1^2-R_2^2)\left(a^2-R_3^2-2R_2^2\ln\frac{a}{R_3}\right)
           \Bigg], \\
  A_{22} &= \frac{1}{16}\Bigg[
           m_2\left(4R_2^2R_3^2-3R_2^4-R_3^4-4R_2^4\ln\frac{R_3}{R_2}\right)
           \notag\\
         &\qquad
           +2(R_2^2-R_3^2)\left(a^2-R_3^2-2R_2^2\ln\frac{a}{R_3}\right)
           \Bigg], \\
  A_{23} &= \frac{1}{16}\left[
           -a^4-R_3^4+2a^2R_2^2+2a^2R_3^2-2R_2^2R_3^2
           -4R_2^2R_3^2\ln\frac{a}{R_3}
           \right], \\
  G_2 &= \frac{1}{16}\Bigg[
	m_2\left(R_2^4+R_3^4+2R_1^2R_2^2-2R_1^2R_3^2-2R_2^2R_3^2
	+4R_1^2R_2^2\ln\frac{R_3}{R_2}\right)
	\notag\\
         &\qquad
           +\left(a^4-R_3^4+2R_1^2R_3^2+2R_2^2R_3^2-2a^2R_1^2-2a^2R_2^2
           +4R_1^2R_2^2\ln\frac{a}{R_3}\right)
           \Bigg].
\end{align}
Finally, for the outermost cumulative flux,
\begin{equation}
  Q^{(3)}=A_{31}p_{1,z}^{(0)}+A_{32}p_{2,z}^{(0)}+A_{33}p_{3,z}^{(0)}+G_3,
\end{equation}
where
\begin{align}
  A_{31} &= \frac{1}{8}(R_1^2-R_2^2)\left(a^2-R_3^2-2R_3^2\ln\frac{a}{R_3}\right), \\
  A_{32} &= \frac{1}{8}(R_2^2-R_3^2)\left(a^2-R_3^2-2R_3^2\ln\frac{a}{R_3}\right), \\
  A_{33} &= \frac{1}{16}\left(
           -a^4-3R_3^4+4a^2R_3^2-4R_3^4\ln\frac{a}{R_3}
           \right), \\
  G_3 &= \frac{1}{16}\left[
	(a^2-R_3^2)(a^2-2R_1^2-R_3^2)+4R_1^2R_3^2\ln\frac{a}{R_3}
	\right].
\end{align}

For convenience, the long-wave evolution equations may therefore be written compactly as
\begin{equation}
  R_{j,t}=\frac{1}{R_j}\partial_z
  \left(
    \sum_{k=1}^3 A_{jk}p_{k,z}^{(0)}+G_j
  \right),\qquad j=1,2,3.
\end{equation}	

\section{Travelling-wave formulation and numerical implementation}

\subsection{Explicit form of the travelling-wave equations}

We consider travelling-wave solutions of the long-wave model in the form
\[
  R_j(z,t)=\mathcal R_j(Z), \qquad Z=z-ct, \qquad j=1,2,3,
\]
where \(c\) is the wave speed. The integrated flux relations
\eqref{eq:TW_integrated_flux} yield
\begin{equation}
  Q^{(j)}+\frac{c}{2}\mathcal R_j^{\,2}=K_j, \qquad j=1,2,3.
\end{equation}
The cumulative fluxes are
\begin{equation}
  Q^{(j)} =
  \sum_{k=1}^3 A_{jk}(\mathcal R_1,\mathcal R_2,\mathcal R_3)\,P_k
  + G_j(\mathcal R_1,\mathcal R_2,\mathcal R_3),
\end{equation}
where \(P_k = p^{(0)}_{k,z}\). The pressure gradients are given by
\begin{align}
  P_1 &=
	\frac{1}{\sigma_1 C}
	\left(
	\frac{\mathcal R_1'}{\mathcal R_1^{2}}
	-\varepsilon^{2}\mathcal R_1'''
	\right),
  \\
  P_2 &=
	P_1
	-
	\frac{1}{\sigma_2 C}
	\left(
	\frac{\mathcal R_2'}{\mathcal R_2^{2}}
	-\varepsilon^{2}\mathcal R_2'''
	\right),
  \\
  P_3 &=
	P_2
	-
	\frac{1}{C}
	\left(
	\frac{\mathcal R_3'}{\mathcal R_3^{2}}
	-\varepsilon^{2}\mathcal R_3'''
	\right).
\end{align}
Substitution yields the coupled system
\begin{equation}
  \sum_{k=1}^3 A_{jk}(\mathcal R)\,P_k
  +
  G_j(\mathcal R)
  +
  \frac{c}{2}\mathcal R_j^{\,2}
  -K_j
  =0,
  \qquad j=1,2,3.
\end{equation}
This constitutes a system of three coupled third-order ordinary differential equations.

\subsection{Matrix formulation}

Collecting the highest-derivative terms gives
\begin{equation}
  \mathsf M(\mathcal R_1,\mathcal R_2,\mathcal R_3)
  \begin{pmatrix}
    \mathcal R_1'''\\
    \mathcal R_2'''\\
    \mathcal R_3'''
  \end{pmatrix}
  =
  \mathsf F(\mathcal R,\mathcal R';c,K_1,K_2,K_3),
\end{equation}
where \(\mathsf M\) is a \(3\times3\) matrix depending on \(A_{jk}\), and
\(\mathsf F\) contains lower-order terms.

Provided \(\mathsf M\) is nonsingular, the system may be written as
\[
  \mathcal R_j''' = f_j(\mathcal R,\mathcal R';c,K_1,K_2,K_3), \qquad j=1,2,3.
\]

\subsection{First-order formulation for continuation}

For numerical continuation, we introduce
\begin{align*}
  y_1&=\mathcal R_1, & y_2&=\mathcal R_1', & y_3&=\mathcal R_1'',\\
  y_4&=\mathcal R_2, & y_5&=\mathcal R_2', & y_6&=\mathcal R_2'',\\
  y_7&=\mathcal R_3, & y_8&=\mathcal R_3', & y_9&=\mathcal R_3''.
\end{align*}
The system becomes
\begin{align}
  y_1' &= y_2, & y_2' &= y_3, & y_3' &= f_1(y;c,K_1,K_2,K_3),\\
  y_4' &= y_5, & y_5' &= y_6, & y_6' &= f_2(y;c,K_1,K_2,K_3),\\
  y_7' &= y_8, & y_8' &= y_9, & y_9' &= f_3(y;c,K_1,K_2,K_3).
\end{align}

\subsection{Boundary conditions and constraints}

The travelling waves are periodic:
\begin{equation}
  y(0)=y(L).
\end{equation}
After rescaling to \(s\in[0,1]\),
\begin{equation}
  y(0)=y(1),
\end{equation}
with \(L\) treated as an unknown. The conservation laws imply constraints on the mean cross-sectional areas:
\begin{align}
  \frac{1}{L}\int_0^L (\mathcal R_2^2-\mathcal R_1^2)\,dZ &= \mathcal V_1,\\
  \frac{1}{L}\int_0^L (\mathcal R_3^2-\mathcal R_2^2)\,dZ &= \mathcal V_2,\\
  \frac{1}{L}\int_0^L (a^2-\mathcal R_3^2)\,dZ &= \mathcal V_3.
\end{align}
Finally, a phase condition is imposed to remove translational invariance, for example
\begin{equation}
  \int_0^1 \langle y(s)-y_{\mathrm{ref}}(s),\, y_{\mathrm{ref}}'(s)\rangle\,ds = 0.
\end{equation}

\subsection{Numerical continuation procedure}

The travelling-wave problem formulated above constitutes a nonlinear
periodic boundary-value problem with integral constraints and unknown
parameters, including the wave speed $c$ and the period $L$.
The system is solved numerically using the continuation software
AUTO~\citep{DoedelAUTO}.

The first-order system is discretised internally by AUTO using
orthogonal collocation on finite elements, and solution branches are
continued using pseudo-arclength continuation. In the computations
presented in §5.3, continuation is performed primarily in the mean
layer thickness parameter $\bar h_1$ with $\bar h_2=\bar h_1$
imposed, while the viscosity ratio $m_2$ is treated as a control
parameter between branch computations.

Limit points along the travelling-wave branches are detected
automatically during continuation and are used to characterise the
termination of the computed solution families.

\end{appen}

\bibliographystyle{jfm}         % jfm.bst comes from the JFM template zip
\bibliography{dh}              % NO .bib extension

\begin{thebibliography}{21}
\expandafter\ifx\csname natexlab\endcsname\relax\def\natexlab#1{#1}\fi
\def\au#1{#1} \def\ed#1{#1} \def\yr#1{#1}\def\at#1{#1}\def\jt#1{\textit{#1}}
  \def\bt#1{#1}\def\bvol#1{\textbf{#1}} \def\vol#1{#1} \def\pg#1{#1}
  \def\publ#1{#1}\def\arxiv#1{#1}\def\org#1{#1}\def\st#1{\textit{#1}}

\bibitem[Chen(1993)]{Chen1993}
{\sc \au{Chen, KangPing}} \yr{1993}  \at{Wave formation in the gravity-driven
  low-{Reynolds} number flow of two liquid films down an inclined plane}.
  \jt{Physics of Fluids}  \bvol{5}~(12),  \pg{3038--3048}.

\bibitem[Doedel {\em et~al.\/}(2007)Doedel, Paffenroth, Champneys, Fairgrieve,
  Kuznetsov, Sandstede \& Wang]{DoedelAUTO}
{\sc \au{Doedel, Eusebius~J.}, \au{Paffenroth, Randy~C.}, \au{Champneys,
  Alan~R.}, \au{Fairgrieve, Thomas~F.}, \au{Kuznetsov, Yuri~A.}, \au{Sandstede,
  Bj{\"o}rn} \& \au{Wang, Xiaonan}} \yr{2007} {\em {AUTO-07P}: Continuation and
  Bifurcation Software for Ordinary Differential Equations\/}.  \publ{Montreal,
  Canada: Concordia University}, available at
  \url{http://indy.cs.concordia.ca/auto/}.

\bibitem[Erken {\em et~al.\/}(2022)Erken, Romanò, Grotberg \&
  Muradoglu]{Erken_Romanò_Grotberg_Muradoglu_2022}
{\sc \au{Erken, O.}, \au{Romanò, F.}, \au{Grotberg, J.B.} \& \au{Muradoglu,
  M.}} \yr{2022}  \at{Capillary instability of a two-layer annular film: an
  airway closure model}.  \jt{Journal of Fluid Mechanics}  \bvol{934},
  \pg{A7}.

\bibitem[Gauglitz \& Radke(1988)]{Gauglitz1988}
{\sc \au{Gauglitz, P.~A.} \& \au{Radke, C.~J.}} \yr{1988}  \at{An extended
  evolution equation for liquid film breakup in cylindrical capillaries}.
  \jt{Chemical Engineering Science}  \bvol{43}~(7),  \pg{1457--1465}.

\bibitem[Gauglitz \& Radke(1990)]{Gauglitz1990}
{\sc \au{Gauglitz, P.~A.} \& \au{Radke, C.~J.}} \yr{1990}  \at{The dynamics of
  liquid film breakup in constricted cylindrical capillaries}.  \jt{Journal of
  Colloid and Interface Science}  \bvol{134}~(1),  \pg{14--40}.

\bibitem[Goren(1962)]{Goren1962}
{\sc \au{Goren, Simon~L.}} \yr{1962}  \at{The instability of an annular thread
  of fluid}.  \jt{Journal of Fluid Mechanics}  \bvol{12}~(2),  \pg{309--319}.

\bibitem[Hammond(1983)]{Hammond1983}
{\sc \au{Hammond, P.~S.}} \yr{1983}  \at{Nonlinear adjustment of a thin annular
  film of viscous fluid surrounding a thread of another within a circular
  cylindrical pipe}.  \jt{Journal of Fluid Mechanics}  \bvol{137},
  \pg{363--384}.

\bibitem[Heil {\em et~al.\/}(2008)Heil, Hazel \& Smith]{Heil2008}
{\sc \au{Heil, Matthias}, \au{Hazel, Andrew~L.} \& \au{Smith, Jaclyn~A.}}
  \yr{2008}  \at{The mechanics of airway closure}.  \jt{Respiratory Physiology
  \& Neurobiology}  \bvol{163}~(1--3),  \pg{214--221}.

\bibitem[Henry {\em et~al.\/}(2014)Henry, Uddin, Thompson, Blyth, Thoroddsen \&
  Marston]{Henry2014}
{\sc \au{Henry, D.}, \au{Uddin, J.}, \au{Thompson, J.}, \au{Blyth, M.~G.},
  \au{Thoroddsen, S.~T.} \& \au{Marston, J.~O.}} \yr{2014}  \at{Multi-layer
  film flow down an inclined plane: experimental investigation}.
  \jt{Experiments in Fluids}  \bvol{55}~(12),  \pg{1859}.

\bibitem[Jiang {\em et~al.\/}(2005)Jiang, Helenbrook, Lin \&
  Weinstein]{Jiang2005}
{\sc \au{Jiang, W.~Y.}, \au{Helenbrook, B.~T.}, \au{Lin, S.~P.} \&
  \au{Weinstein, S.~J.}} \yr{2005}  \at{Low-{R}eynolds-number instabilities in
  three-layer flow down an inclined wall}.  \jt{Journal of Fluid Mechanics}
  \bvol{539},  \pg{387--416}.

\bibitem[Joseph {\em et~al.\/}(1997)Joseph, Bai, Chen \& Renardy]{Joseph1997}
{\sc \au{Joseph, D.~D.}, \au{Bai, R.}, \au{Chen, K.~P.} \& \au{Renardy, Y.~Y.}}
  \yr{1997}  \at{Core-annular flows}.  \jt{Annual Review of Fluid Mechanics}
  \bvol{29},  \pg{65--90}.

\bibitem[Kao(1965{\natexlab{{\em a\/}}})]{Kao1965b}
{\sc \au{Kao, T.~W.}} \yr{1965{\natexlab{{\em a\/}}}}  \at{Role of the
  interface in the stability of stratified flow down an inclined plane}.
  \jt{Physics of Fluids}  \bvol{8}~(12),  \pg{2190--2194}.

\bibitem[Kao(1965{\natexlab{{\em b\/}}})]{Kao1965a}
{\sc \au{Kao, T.~W.}} \yr{1965{\natexlab{{\em b\/}}}}  \at{Stability of
  two-layer viscous stratified flow down an inclined plane}.  \jt{Physics of
  Fluids}  \bvol{8}~(5),  \pg{812--820}.

\bibitem[Lamnawar {\em et~al.\/}(2013)Lamnawar, Zhang \& Maazouz]{Lamnawar2013}
{\sc \au{Lamnawar, K.}, \au{Zhang, H.} \& \au{Maazouz, A.}} \yr{2013}
  \at{Coextrusion of multilayer structures, interfacial phenomena}.  \bt{In
  {\em Encyclopedia of Polymer Science and Technology\/}}.  \publ{Wiley}.

\bibitem[Loewenherz \& Lawrence(1989)]{Loewenherz1989}
{\sc \au{Loewenherz, D.~S.} \& \au{Lawrence, C.~J.}} \yr{1989}  \at{The effect
  of viscosity stratification on the stability of a free surface flow at low
  {Reynolds} number}.  \jt{Physics of Fluids A: Fluid Dynamics}  \bvol{1}~(10),
   \pg{1686--1693}.

\bibitem[Ogrosky(2021)]{Ogrosky2021}
{\sc \au{Ogrosky, H.~Reed}} \yr{2021}  \at{Impact of viscosity ratio on falling
  two-layer viscous film flow inside a tube}.  \jt{Physical Review Fluids}
  \bvol{6}~(10).

\bibitem[Peng {\em et~al.\/}(2025)Peng, Feng, Xue, Zhou, Wang, Zhong \&
  Ku]{Peng2025}
{\sc \au{Peng, Jianjun}, \au{Feng, Run}, \au{Xue, Meng}, \au{Zhou, Erhao},
  \au{Wang, Junhua}, \au{Zhong, Zhidan} \& \au{Ku, Xiangchen}} \yr{2025}
  \at{Research progress and engineering applications of viscous fluid
  mechanics}.  \jt{Applied Sciences}  \bvol{15}~(1),  \pg{357}.

\bibitem[Rayleigh(1878)]{Rayleigh1878}
{\sc \au{Rayleigh, Lord}} \yr{1878}  \at{On the instability of jets}.
  \jt{Proceedings of the London Mathematical Society}  \bvol{s1-10}~(1),
  \pg{4--13}.

\bibitem[Roman{\`o}(2026)]{romano2026human}
{\sc \au{Roman{\`o}, Francesco}} \yr{2026}  \at{Human lungs fluid mechanics: an
  overview of current modelling techniques}.  \jt{The European Physical Journal
  E}  \bvol{49}~(5),  \pg{38}.

\bibitem[Tomotika(1935)]{Tomotika1935}
{\sc \au{Tomotika, S.}} \yr{1935}  \at{On the instability of a cylindrical
  thread of a viscous liquid surrounded by another viscous fluid}.
  \jt{Proceedings of the Royal Society of London. Series~{A}, Mathematical and
  Physical Sciences}  \bvol{150}~(870),  \pg{322--337}.

\bibitem[Traore(2024)]{Traore2024}
{\sc \au{Traore, Awa}} \yr{2024}  \at{Dynamics of a multi-fluid system:
  Parallel study of three fluids with a passive core inside a cylinder and
  three fluids inside a pair of concentric cylinders}. PhD thesis, The
  University of Alabama.

\end{thebibliography}
\end{document}